\documentclass[12pt,english]{article}

\usepackage[utf8]{inputenc}
\usepackage[T1]{fontenc}
\usepackage[english]{babel}

\usepackage{newtxtext,newtxmath}  % Cleaner than txfonts
\usepackage{microtype}            % Subtle kerning and spacing improvements

\usepackage{geometry}
\usepackage{setspace}
\usepackage{ragged2e}
\justifying

\usepackage[medium,compact]{titlesec}
\titleformat{\section}{\large\bfseries}{\thesection}{1em}{}
\titleformat{\subsection}{\normalsize\itshape}{\thesubsection}{1em}{}

\usepackage[style=authoryear,natbib,sorting=none]{biblatex}
\usepackage[hidelinks]{hyperref}
\hypersetup{
  colorlinks=true,
  linkcolor=blue,
  urlcolor=cyan,
  citecolor=blue
}

\usepackage{graphicx}
\usepackage{float}
\usepackage{wrapfig}
\graphicspath{{graphics}}

\usepackage{mathtools, bm, cancel, physics, dsfont,array}

\usepackage[font=small,labelfont=bf]{caption}

\usepackage[bottom]{footmisc}

\usepackage{tikz}
\usetikzlibrary{decorations.pathreplacing, bending, arrows.meta,automata}

\usepackage[toc,page]{appendix}

\usepackage{authblk}

\usepackage{xcolor}
\usepackage{xspace}
\usepackage{ifthen}
\usepackage{lipsum}
\usepackage{tablefootnote}

\usepackage{enumitem}
\setlist[itemize]{itemsep=0.2em, topsep=0.2em}

\newcommand{\comment}[1]{}

\newcommand{\showcomments}{true}

\newcommand{\gui}[1]%
{\ifthenelse{\equal{\showcomments}{true}}%
{{\color{magenta}{\small [\textbf{gui:} #1}]}}{\xspace}}%

\newcommand{\remove}[1]%
{\ifthenelse{\equal{\showcomments}{true}}%
{{\color{brown}{\small [\textbf{remove:} #1}]}}{\xspace}}%

\newcommand{\richard}[1]%
{\ifthenelse{\equal{\showcomments}{true}}%
{{\color{blue}{\small [\textbf{Richard:} #1}]}}{\xspace}}%

\newcommand{\Hcal}{\mathcal{H}}
\newcommand{\Tcal}{\mathcal{T}}
\newcommand{\Acal}{\mathcal{A}}

\newcommand{\Scal}{\mathcal{S}}
\newcommand{\Ocal}{\mathcal{O}}

\usepackage{amssymb, amsthm}
\newtheorem{theorem}{Theorem}[section]
\theoremstyle{definition}
\newtheorem{definition}[theorem]{Definition}

\titleformat*{\section}{\bf}
\titleformat*{\subsection}{\it}
\titleformat*{\subsubsection}{\it}

\begin{document}

\pagenumbering{roman}

\begin{titlepage}

\baselineskip=15.5pt \thispagestyle{empty}

% indentation
\setlength\parindent{12pt}%https://www.overleaf.com/project/60dc48a3385d18d0ccb2e7cb

\begin{center}
    {\fontsize{19}{24}\selectfont \bfseries Realism and Ontology 
    \vspace{0.1cm} \\ in Quantum Mechanics and String Theory}
\end{center}

\vspace{0.1cm}

\renewcommand*{\thefootnote}{\fnsymbol{footnote}}

\begin{center}
     Richard Dawid$^{1}$, \footnote{richard.dawid@philosophy.su.se
}   Guilherme Franzmann$^{1,2,3}$ \footnote{guilherme.franzmann@su.se}
\end{center}

\renewcommand*{\thefootnote}{\arabic{footnote}}
\setcounter{footnote}{1}

\vspace{-0.3cm}
\begin{center}
\rule{14cm}{0.08cm}
\end{center}

% \begin{center}
\footnotesize
% \vskip2pt
   \noindent
    % \vskip4pt
   \noindent 
    \textsl{$^1$Department of Philosophy, Stockholm University, Stockholm, Sweden}

    \vskip4pt
   \noindent 
    \textsl{$^2$Nordita, KTH Royal Institute of Technology and Stockholm University,
Hannes Alfvéns v\"ag 12, SE-106 91 Stockholm, Sweden}

    \vskip4pt
   \noindent 
    \textsl{$^3$Basic Research Community for Physics e.V., Mariannenstraße 89, Leipzig, Germany}

\normalsize

% \end{center}

\vspace{0.5cm}

%%% ABSTRACT %%%
\hrule
\vspace{0.3cm}

\thispagestyle{empty}

\begin{abstract}
Dualities in physics have challenged traditional forms of scientific realism by undermining the idea that theories describe a unique underlying ontology. In this paper, we develop a new perspective on scientific realism that responds to this challenge. We argue that while realist commitment remains appropriate at the level of a theory’s full formal structure, ontological commitment should be treated as tied to specific empirical contexts rather than to a fixed, real ontology. Our proposal draws inspiration from Dennett’s notion of a  ``compression algorithm'' as a defining criterion of a scientific theory. On this basis, we separate realism from ontological commitment. To clarify the stakes of this distinction, we contrast our approach with common core realism, which locates ontology in the invariant structure shared by dual models. Focusing on dualities in quantum mechanics and string theory, we show how our view accommodates ontological pluralism while preserving a robust form of structural realism.
\begin{center}\textbf{Word count:} 15884 \end{center}
\end{abstract}

\vskip10pt
\hrule
\vskip10pt

\newpage
\setlength{\topmargin}{0in}
\tableofcontents

\end{titlepage}

\clearpage

\pagenumbering{arabic}
\setcounter{page}{1}

\section{Introduction}

Dualities are one of the most conspicuous characteristics of fundamental physics. They play a pivotal role in string theory and show up in various contexts of quantum field theory. In a slightly broader sense, quantum mechanics itself is characterized by dual representations.

One of the central challenges dualities pose for scientific realism is the apparent decoupling between a theory’s fundamental structure and its ontological content. Dual models often differ dramatically in their ontology--featuring different fundamental entities, interactions, and even different spacetime structures--while remaining empirically equivalent. This has led to proposals that may be subsumed under the name \textit{common core realism} (\cite{Matsubara2018-MATSIS-4}, \cite{Rickles2017-RICDTS}, \cite{Huggett2017-HUGTS-2}), and more recently \citep{deHaroButterfield}, which seek to identify invariant structure across dual models as the proper target of both realist and ontological commitment. 

In this paper, we propose a different approach. We argue that realist commitment regarding a theory that has different dual representations should be decoupled from the endorsement of a specific ontology.
Our proposal is related to and bears some resemblance to \cite{Dennett1991-DENRP}'s defense of ``real patterns''. There, Dennett argues that realist commitment should not depend on whether a structure is fundamental, but on whether it supports a successful compression algorithm, i.e., whether it yields explanatory and predictive power.
\textit{Dennett's proposal amounts to separating realism from fundamentalism}: one can be a realist about a pattern without insisting that it reflects the ultimate building blocks of reality. Dennett’s point is that explanatory success, not metaphysical depth, should guide our realism. 
We adopt Dennett's concept of a compression algorithm but deploy it in a slightly different way: whereas Dennett’s account uses it to separate realism from fundamentalism, our use of it serves to separate realism from ontology.

A scientific theory, on our account, amounts to a compression algorithm in Dennett's sense. Realist commitment, on this account, expresses trust in the theory's viability (empirical adequacy) in a given empirical domain or, if one endorses a universal final theory claim about a given theory, with regard to the entire world.
But realist commitment does not imply the endorsement of an ontology of the world. 

Although ontology is thus separated from scientific realism, it still plays an important role in the understanding of the world. It is tied to the specific contingent characteristics of the theory’s dynamics, degrees of freedom, solutions, and ground state that provide the basis for the human observer’s empirical access to the world. Thereby, it points at a specific observational perspective on the world, representing the theory in terms of human intuitions about elementary objects that are rooted in that perspective. In other words, we endorse a view of ontology that is, in a deep sense, perspectival \citep{Adlam_2024}, sensitive to the empirical interface between theory and world.  
 We will call the described take on ontology \textit{observer-based ontology}.

To avoid ambiguities in what follows, it is helpful to introduce a few clarifying distinctions between our terminology and nearby positions in the literature. In our account,  \emph{realism} refers to commitment to the approximate truth of a theory’s full formal structure, including the invariant relations preserved across dual models. By contrast, \emph{ontology} designates the entities or structures that arise only in regimes where the theory affords empirical access, such as subsystem decompositions in quantum mechanics or near-classical limits in string theory. Because these regimes exist only relative to the physical presence of observers, ontology in our sense is \emph{observer-based}. Crucially, this does \emph{not} mean that ontology is observation-dependent (varying with specific measurement outcomes), nor perspectivalist in the sense of \cite{Massimi2022-MASPRB}, because the human perspective, and therefore the adequate choice of ontology, is enforced by the universe's dynamics and the place in the universe we happen to be in. Extracting the ontology, therefore, is not a matter of pragmatic choice. Realism thus targets the truth of the theory’s observer-independent invariant structure, while ontology reflects the structures that become salient only when empirical access is physically realized.

The described separation between realist commitment and ontology allows us to retain the epistemic virtues of scientific realism even in the presence of dualities, without committing to a unique or real ontology. It also provides a basis for distinguishing between two equally important ways of speaking about the world: an observer-independent commitment to the (approximate) truth of a theory can be expressed in the form of realist commitment, and a conception of the world in terms of the empirical perspective we happen to have on the world that can be expressed by spelling out an ontology.

In Sec. \ref{sec:dualities}, we will give a brief introduction to dualities in quantum mechanics and string theory. Sec. \ref{sec:common_core} will then discuss common core realism. We will base our discussion on the detailed formal analysis of the common core in \citep{deHaroButterfield}.
In Sec. \ref{sec:issuesCCR}, we discuss problems of common core realism, which then leads to the presentation of our proposal in Sec. \ref{sec:proposal1}, and its application to quantum mechanics in Sec. \ref{sec:proposal2} and to string theory in Sec. \ref{sec:new_ontology_ST}. 
   
\section{Dualities}\label{sec:dualities}

Dualities have become a central feature of modern theoretical physics, challenging the assumption that a physical system must admit a unique mathematical formulation. They reveal cases where distinct theoretical descriptions, often involving different ontologies, yield the same empirical predictions. In this section, we introduce dualities in two key domains: quantum mechanics and string theory. Our goal is not only to clarify the structure and function of dualities in each case, but also to lay the groundwork for questioning the ontological significance of the structures involved. We begin with quantum mechanics, where dualities emerge from different choices of tensor product structure on the Hilbert space, and then turn to string theory, where dualities connect distinct near-classical limits of the theory despite its full formulation remaining elusive.

\subsection{Dualities in Quantum Mechanics}\label{sec:dualities_qm}

A quantum system\footnote{We will exclusively consider standard quantum-mechanical systems described on a separable Hilbert space, without invoking the algebraic and locality structures characteristic of quantum field theory. We adopt however the Heisenberg picture, which allows our considerations to extend naturally to field-theoretic contexts and to parallel the string-theory case.} $\mathcal{Q}$ is typically represented by a vector state, $\ket{\Psi}$, in a Hilbert space, $\Hcal$, together with the algebra of operators acting on it, $\mathcal{A} = \mathcal{B}(\Hcal)$. The Hermitian operators $\Ocal = \{ \hat{\Ocal} \in \Acal \,|\, \hat{\Ocal}^\dagger = \hat{\Ocal} \}$, define the observables of the system. Although $\Ocal$ is not closed under multiplication, it is sufficient to span $\Acal$: every operator $\hat{A} \in \Acal$ can be written uniquely as a complex linear combination of Hermitian operators. Since Hilbert spaces of the same dimensionality are isomorphic and $\mathcal{A}$ spans all linear maps on $\Hcal$, it is the specific representation of observables—particularly the choice of Hamiltonian $\hat{H} \in \Ocal$, together with additional algebraic or symmetry structures in $\mathcal{A}$—that distinguishes different quantum systems. In the Heisenberg picture, the time evolution of observables is given by the equation
\begin{equation}
\frac{d}{dt} \hat{\Ocal} = i [\hat{H}, \hat{\Ocal}] \,. 
\end{equation}

Among the eigenstates of the Hamiltonian, we identify the vacuum state, $\ket{0}$, as the state of lowest energy. 
Moreover, since $\mathcal{A}=\mathcal{B}(\Hcal)$ acts irreducibly on $\Hcal$, any nonzero vector is cyclic; in particular, any other state can be obtained from $\ket{0}$ by the action of an operator in $\mathcal{A}$ (we say the vacuum state is \textit{cyclic} in Hilbert space). 
Therefore, once a state-preparation prescription is fixed, the empirical predictions of $\mathcal{Q}$ can be encoded in vacuum expectation values of observables (and their conjugations by preparation operators),
\begin{equation}
    \langle \hat{\Ocal} \rangle :=\bra{0}  \hat{\Ocal} \ket{0} \,.\label{eq:QM_empirical_import}
\end{equation}
Therefore, following \cite{deHaroButterfield}, we can succinctly identify the quantum system $\mathcal{Q}$ with the triple composed by the space of states, $\Hcal$, the space of observables, $\Ocal$, and the dynamical evolution represented by the Hamiltonian, $\hat H$, 
\begin{equation}
\mathcal{Q} := (\Hcal,\Ocal, \hat{H}) \longrightarrow (\ket{0}, \Ocal, \hat{H}) \,,
\end{equation}
where $\ket{0}$ serves as a fixed reference state in the Heisenberg picture, allowing us to package state-dependence into the choice of operators acting on $\ket{0}$\footnote{In particular, for any $\ket{\psi}\in\Hcal$ there exists $\hat A\in\mathcal{A}$ such that $\ket{\psi}\propto \hat A\ket{0}$, and hence $\bra{\psi}\hat{\Ocal}\ket{\psi}\propto \bra{0}\hat A^\dagger \hat{\Ocal}\hat A\ket{0}$.}.

\subsubsection{Composition of Quantum Systems}

For all practical purposes, we tend to think of the experimental arrangements in our labs as a composite of different quantum systems\footnote{For a more extensive qualitative discussion of independent quantum systems, see \citep{Franzmann:2024rzj}. A more formal treatment can be found in \citep{Summers:2008zza,EmerGe_proj_2}.}, $\mathcal{Q}_i$. Therefore,  we need to define the rules of composition of such systems so that the composite system is also a quantum system as introduced above. This is done by introducing a \textit{tensor product structure} (TPS):

\vspace{0.2cm}
\begin{definition}[TPS]
    A TPS $\Tcal$ of Hilbert space $\Hcal$ is an equivalence class of isomorphisms $t:\Hcal \rightarrow \otimes_i \Hcal_i,$ where $t_1 \sim t_2$ whenever $t_1 t_2^{-1}$ can be written as a product of local unitaries $\otimes_i U_i$ and permutations of subsystems \citep{Cotler:2017abq}. \label{def:TPS}
\end{definition}

\vspace{0.2cm}

\noindent Therefore, the composite (aka global) system's Hilbert space is 
\begin{equation}
    \bigotimes_i \Hcal_i \xlongrightarrow[]{t^{-1}} \mathcal{H} \,,
\end{equation}
where $t^{-1}$ is the inverse of the isomorphism map. Moreover, given chosen reference states $\ket{0}_i\in\Hcal_i$, the corresponding product state is written as
\begin{equation}
    \bigotimes_i \ket{0}_i \xlongrightarrow[]{t^{-1}} \ket{0} \in \Hcal,
\end{equation}
which coincides with the global ground state only in special cases, e.g.\ for dynamically independent subsystems.

However, this composition rule does not straightforwardly apply to the space of operators acting on the newly formed global Hilbert space. Arguably, the most surprising aspect of quantum mechanics is that the set of \emph{simple} tensor-product (``local'') operators is strictly smaller than the full operator algebra on the composite system,
\begin{equation}\label{eq:sep_operators}
     \bigotimes_i \hat{A}_{i} \xlongrightarrow[]{t^{-1}}  \hat{A}_s \,, \text{s.t. } \{\hat{A}_s\} \subset \mathcal{A}\,,
\end{equation}
where $\hat{A}_i \in \mathcal{A}_i$ and $\hat{A}_s \in \mathcal{A}$. We use the subscript $s$ to indicate operators that are \emph{separable/product} with respect to the chosen TPS. Naturally, there exist global operators that cannot be written in the product form \eqref{eq:sep_operators}; these include operations that can generate entanglement from initially separable states (in the Schrödinger picture). More generally, such operators can be expressed as sums of product terms,
\begin{equation}
    \sum_n c_n \bigotimes_i \hat{A}_{i}^{(n)}\,,
\end{equation}
with at least two nonvanishing coefficients, and which in general cannot be rewritten as a single product operator.

Due to the existence of genuinely non-product (entangling) global operators, the notion of a composite system is nontrivial in quantum mechanics,  
\begin{equation}
    ``\bigotimes_i \mathcal{Q}_i \,\,  \cancel{\xlongrightarrow[]{t^{-1}}} \,\, 
 \mathcal{Q} \,, ''
\end{equation}
and the naive expectation is not realized: assembling systems yields new global states and operations not reducible to independent local ones. Thus, let's consider the reverse process of decomposing a system into parts instead.

\subsubsection{Breaking Down Quantum Systems}

Now we start with a quantum system $\mathcal{Q}$ and consider a map $t$ on its Hilbert space such that
\begin{equation}
    t: \Hcal \rightarrow \bigotimes_i \Hcal_i\,,
\end{equation}
factorizing the global Hilbert space, where each factor is associated with a local system. Thus, the choice of this map endows the global Hilbert space with a notion of locality, allowing us to talk about local operations: an operator $\hat{A}\in \mathcal{A}$ is local to subsystem $i$ if $t\hat{A} t^{-1}$ is local to $\Hcal_i$ \citep{Cotler:2017abq}.

At first glance, shifting focus from \textit{composing} to \textit{decomposing} quantum systems may not seem especially illuminating. But there is a crucial asymmetry: while the composition of subsystems naturally introduces a TPS on the global Hilbert space, decomposing a system does not come with a canonical choice of TPS. When we compose subsystems, we preserve the original notion of parts and their local structure. In contrast, when we start with a global system and seek to break it into subsystems, there is no \textit{a priori} notion of what the parts should be. As a result, there exist multiple \textit{distinct} TPSs, each leading to a different notion of subsystem structure\footnote{Here, ``distinct'' refers to TPSs that are different as mathematical decompositions of the Hilbert space, independently of whether they are related by a unitary transformation that preserves the Hamiltonian. In the literature, following \cite{Cotler:2017abq}, the relevant notion of \emph{equivalence} is instead Hamiltonian-relative: two TPSs are considered equivalent when related by a global unitary $U$ such that $U\hat{H}U^{\dagger}=\hat{H}$, up to local unitaries and permutations of factors. These notions have been carefully disentangled in \citep{SoulasFranzmannDiBiagio}, where it's also shown that, while $\hat{H}$ by itself cannot fix a unique TPS, the combined data $(\hat{H},|\psi\rangle)$ may suffice to select one in a relational and physically meaningful sense \citep{Stoica:2021rqi}.\label{ft:distinct_inequi}}.

As the TPS is an equivalence class of isomorphism maps (effectively picking up a set of bases), it preserves the spectra of all the global operators acting on the global Hilbert space. However, different TPSs break down some of these operators into separable ones, which also preserves the spectra. The choice of TPS will determine which operators are broken down into separable ones. 

To make the discussion concrete, let's consider the following simple quantum system, 
\begin{align}
    \Hcal &= \mathbb{C}^4\,, \\
    \hat{H} & = \begin{pmatrix} 
J & h & h & 0 \\ 
h & -J & 0 & h \\ 
h & 0 & -J & h \\ 
0 & h & h & J \\ 
\end{pmatrix}\, \text{, with eigenvalues } \{E_n\}= \pm\{J, \sqrt{J^2 + h^2}\}\,.
\end{align}
Looking back at the definition of TPS, this description instantiates the trivial TPS, $\mathcal{T}_0$, where the system is not partitioned at all. Therefore, the Hamiltonian is simply the free Hamiltonian of the system, and a change of basis can diagonalize it to a linear sum of its eigenvalues multiplied by projections that form a basis for diagonal matrices (in the language of states, they correspond to the density operators associated with the four different energy levels). 

 A less trivial TPS, $\mathcal{T}_1$, is the following:
 \begin{equation}
 \begin{aligned}
    \Hcal   = \mathbb{C}^4 & \xlongrightarrow[]{t_1} \Hcal_1 \otimes \Hcal_2 = \mathbb{C}^2 \otimes \mathbb{C}^2    \\
    \hat{H}  = \begin{pmatrix} 
J & h & h & 0 \\ 
h & -J & 0 & h \\ 
h & 0 & -J & h \\ 
0 & h & h & J \\ 
\end{pmatrix} & \xlongrightarrow[]{t_1} J
\begin{pmatrix} 
1 & 0 \\ 
0 & -1 \\ 
\end{pmatrix} \otimes
\begin{pmatrix} 
1 & 0 \\ 
0 & -1 \\ 
\end{pmatrix} + h
\left[
\begin{pmatrix} 
0 & 1 \\ 
1 & 0 \\ 
\end{pmatrix} \otimes 
\begin{pmatrix} 
1 & 0 \\ 
0 & 1 \\ 
\end{pmatrix} + \begin{pmatrix} 
1 & 0 \\ 
0 & 1 \\ 
\end{pmatrix} \otimes \begin{pmatrix} 
0 & 1 \\ 
1 & 0 \\ 
\end{pmatrix}  \right]  \\
& = J \sigma^z_1 \sigma^z_2 + h (\sigma^x_1  + \sigma^x_2)\,, 
\end{aligned}
\end{equation}
which factorizes the global Hilbert space as the composition of two qubits interacting with strength $J$ under a magnetic field $h$ in the $x$-direction. 

At this point, we ask ourselves: what really is the physical system? A free $4$-level system or an interacting $2$-qubit system in a magnetic field? At this point, we simply cannot tell them apart. The descriptions are \textit{dual}: 

\begin{definition}[Duality]
A duality is an \emph{isomorphism} between two theories\footnote{We will expand on the notion of a theory in Sec. \ref{sec:common_core} where we review common core realism.} \citep{deHaroButterfield}. \label{duality1}
\end{definition}

\noindent Since the two TPSs above are related by an isomorphism, this example presents a duality. Crucially, it preserves the spectrum of the Hamiltonian, but also allows the identification of observables across each description\footnote{Note that in \cite{deHaroButterfield} their emphasis is on states, instead.}.

One could contest the tension raised so far about the \textit{true reality} of the physical system as artificial, given the definition of duality being too loose. In fact, some authors \citep{Cotler:2017abq} invoke a stricter definition of dualities in quantum mechanics, which demands additional structures to be invariant under the isomorphism. To illustrate that, consider the same system again, but now under a yet different TPS, $\mathcal{T}_2$:
\begin{equation}
 \begin{aligned}
    \Hcal = \mathbb{C}^4 & \xlongrightarrow[]{t_2} \Hcal_1' \otimes \Hcal_2' = \mathbb{C}^2 \otimes \mathbb{C}^2   \,, \\
    \hat{H} = \begin{pmatrix} 
2h & 0 & J & 0 \\ 
0 & 0 & 0 & J \\ 
J & 0 & -2h & 0 \\ 
0 & J & 0 & 0 \\ 
\end{pmatrix} & \xlongrightarrow[]{t_2} J
\begin{pmatrix} 
0 & 1 \\ 
1 & 0 \\ 
\end{pmatrix} \otimes
\begin{pmatrix} 
1 & 0 \\ 
0 & 1 \\ 
\end{pmatrix} + h
\left[
\begin{pmatrix} 
1 & 0 \\ 
0 & -1 \\ 
\end{pmatrix} \otimes 
\begin{pmatrix} 
1 & 0 \\ 
0 & -1 \\ 
\end{pmatrix} + \begin{pmatrix} 
1 & 0 \\ 
0 & -1 \\ 
\end{pmatrix} \otimes \begin{pmatrix} 
1 & 0 \\ 
0 & 1 \\ 
\end{pmatrix}  \right]\, \,, \\
& = h \mu^z_{1} \mu^z_{2} + J \mu^x_{1}  + h \mu^z_{1}\,, 
\end{aligned}
\end{equation}
which factorizes the global Hilbert space as the composition of two qubits interacting with strength $h$, with the qubit-$1$ under a magnetic field $J$ in the $x$-direction and another $h$ on the $z$-direction. Despite having a $2$-qubit system again, these are not the same qubits as before, which we indicated by the ' in $\Hcal_i'$ above. 

This example can be easily generalized for $N$-qubits, where the two Hamiltonians for each TPS are equal and written as: 
\begin{equation}
    H = J\sum_{i=1}^{N-1} \sigma_i^z \sigma_{i+1}^z + h\sum_{i=1}^N \sigma_i^z = J\sum_{i=1}^{N} \mu^x_i +  h\sum_{i=1}^{N-1} \mu_i^z \mu_{i+1}^z - J\mu_N^x + h\mu_1^z\,.
\end{equation}
One can easily check that $\mathcal{T}_1$ and $\mathcal{T}_2$ are not equivalent, as the two Hamiltonians can be mapped by the following transformation:
\begin{equation}
    \mu_i^z = \prod_{j\leq i} \sigma^x_j\,,\quad \mu^x_i = \sigma^z_i \sigma_{i+1}^z\,, \quad \mu_N^x=\sigma^z_N\,, 
\end{equation}
which is not a product of local unitary transformations across each subsystem's operators, nor a permutation of them (see Definition \ref{duality1}). 

The crucial difference between the comparison of $\mathcal{T}_0$ and $\mathcal{T}_1$, and of $\mathcal{T}_1$ and $\mathcal{T}_2$ is that the Hamiltonian under the latter preserves the same degree of local interactions, i.e., a maximum of two systems interacting at the time. This motivates the  stricter definition of duality: 

\begin{definition}[k-Duality]
    Two TPSs are dual if they are not equivalent and their associated Hamiltonians, which have the same spectra, are $k$-local \citep{Cotler:2017abq}\footnote{Note that $k$-locality is a very general and useful definition, extending itself even to QFTs: typically we do not see interacting terms of arbitrary orders in the Standard Model, but restricted to quartic interactions, which is tightly connecting to their being renormalizable.}. \label{kduality} 
\end{definition}

\noindent A Hamiltonian is \textit{$k$-local} if it can be written as a sum of terms, each of which acts non-trivially on at most $k$ subsystems. In our context, this means the interaction structure involves at most $k$-body couplings. For example, a 2-local Hamiltonian involves only pairwise interactions between subsystems.

Note that the existence of multiple TPSs with the same local structure in the Hamiltonian heightens the tension around the question of what the physical system really consists of\footnote{There is a sense in which our analysis has been naive so far, especially in the context of the Heisenberg picture where the very Hamiltonian evolution defines an orbit of distinct TPSs \citep{Stoica:2021rqi,SoulasFranzmannDiBiagio} (but equivalent ones a la \cite{Cotler:2017abq}). For the rest of the paper, one can simply think of the \textit{kinematically-induced} TPS given by the system's decomposition at some initial time \citep{Zanardi:2004zz}. However, the truly fundamental notion of subsystems is encapsulated by the algebraic structure defined on a Hilbert space, in terms of its subalgebras and their relationships, which is best encapsulated in the language of descriptors \citep{Deutsch:1999jb,B_dard_2021}. We briefly review that in the Appendix \ref{sec:descriptors}.}. Before we turn to our proposed solution to this challenge in Sec. \ref{sec:proposal}, we will first introduce the notion of dualities in the context of string theory, and then spell out the difficulties with the proposed solution to the problem of reality in the presence of dualities given by common core realism discussed in \citep{deHaroButterfield}.

\subsection{Dualities in String Theory}\label{sec:dualities_ST}

Duality relations play a crucial role in the overall understanding of string theory. We will first discuss S- and T-dualities that link the various types of string theory and then briefly look at the AdS/CFT correspondence that links string theory to a conformal field theory.

\subsubsection{S- and T-duality}

To understand the status of S- and T-duality \citep{polchinski2015dualitiesfieldsstrings}, one needs to say a few words about the way string theories are constructed. String theory, at any rate at the current level of understanding, lacks a full formal formulation. The starting point for developing a string theory is provided by perturbation theory: individual strings interact with each other, which, at an effective level, can be represented in terms of a perturbative description up to some orders of the effective string coupling. The effective interaction structure satisfies a certain gauge symmetry. Several different boundary conditions can be set for the oscillating string, which provides the basis for constructing five different types of superstring theory (string theory with fermionic degrees of freedom).
Each of those types of string theory, while constructed on a perturbative basis, is understood to represent and determine a full theory that is not constrained to the low coupling regime and also includes non-perturbative effects. For small string couplings, those non-perturbative effects remain small. Once one approaches the strongly coupled regime, non-perturbative effects take over, and perturbative string theory becomes unreliable\footnote{Note, however, that the specification of the string coupling is itself controlled by the string dynamics. Thus, assuming a small string coupling already in itself forecloses a fundamental understanding of the dynamics of string theory.}.

A type of string theory's gauge structure is implied by the boundary conditions set for the oscillating string and cannot be freely chosen. Importantly, the sets of boundary conditions that correspond to the 5 types of superstring theory lead to various different gauge structures. In other words, the interaction structures of the various types of string theory look substantially different from a perturbative point of view.

This is where S-and T-duality come in. They suggest that the 5 types of string theory are empirically equivalent once one takes non-perturbative effects into account. 

\paragraph{T-duality.} It relates a string theory with a given radius $R$ of a compactified dimension to a different string theory with radius $1/R$ (in units of the string length $l_s$). 

\noindent Closed strings moving along the compact dimension are characterized by two discrete numbers: the quantized momentum along the compact dimension (called the Kaluza-Klein state) and the times the string is wrapped around the compact dimension (called the winding number). The Kaluza-Klein state of the initial theory is equal to the winding number of the T-dual and vice versa. Open strings moving along the compact dimension have no winding number. Under T-duality, however, the boundary conditions along the dualized circle are exchanged: Neumann boundary conditions become Dirichlet boundary conditions (and vice versa). As a result, open-string endpoints become localized in the dual circle, so that open strings are naturally described as ending on hypersurfaces (D-branes) that are extended in the non-compact directions\footnote{A brane is a generalization of a membrane to an arbitrary number of spatial dimensions. A D-brane (short for Dirichlet brane) is a dynamical object on which open strings can end, with Dirichlet boundary conditions fixing their endpoints.}. Therefore, D-branes emerge as a necessary ingredient of the full non-perturbative string framework.

\paragraph {S-duality.} It relates a string theory with string coupling $g_s$ to a string theory with coupling $1/g_s$, mapping the weakly coupled (perturbative) regime of one theory to the strongly coupled (non-perturbative) regime of the other. 

Both kinds of duality connect a near-classical limit in one type of string theory (where the perturbative description works well) to a dual description in terms of a deep quantum situation in a different type of string theory. In the case of S-duality, this is straightforward as the duality connects a weakly-coupled theory to a strongly-coupled one. In the case of T-duality, we have a near-classical scenario if the compact spatial dimension is large: in that case, the spectrum of transversal momenta is near-continuous and winding modes are so heavy that they play no role. In other words, the situation is locally well described by the movement of a localizable string in an open dimension. This scenario is linked to a dual description on a very small dimension, where the energy gaps between transversal momenta are big and lots of light winding-modes enter the picture. Represented in terms of the dual type of string theory, the situation thus does not look like the movement of localizable strings in a macroscopically extended space at all.    

\begin{figure}
\centering
\scalebox{1.0}{
\begin{tikzpicture}[-, auto, node distance=2.5cm, semithick]
%\tikzstyle{every state}=[text=black, minimum size=18mm]

\node(A) {$SO(32)$};
\node(B) [left of=A]{$E_{8}\times E_{8}$};
\node(C) [above left of=B]{$M$};
\node(D) [below left of=C]{$IIA$};
\node(E) [right of=A]{$I$};
\node(F) [left of=D]{$IIB$};

\path (A) edge node{\textbf{S}} (E);
\path (A) edge node{\textbf{T}} (B);
\path (B) edge node{} (C);
\path (F) edge node{\textbf{T}} (D);
\path (D) edge node{} (C);

\path (F) edge [loop left, looseness=15, min distance=20mm, out=135, in=225] node[left]{\textbf{S}} (F);
\end{tikzpicture}}
\vspace{-1.5cm}
\caption{Dualities between the five types of Superstring theories (type IIA, type IIB, heterotic $E_{8}\times E_{8}$, heterotic $SO(32)$, and type I) and M-theory. The extended 11th dimension of M-theory corresponds to the strength of the string coupling of type IIA and heterotic $E_{8}\times E_{8}$ in the dual pictures. Type IIB is S-self-dual.}\label{fig2} 
\end{figure}
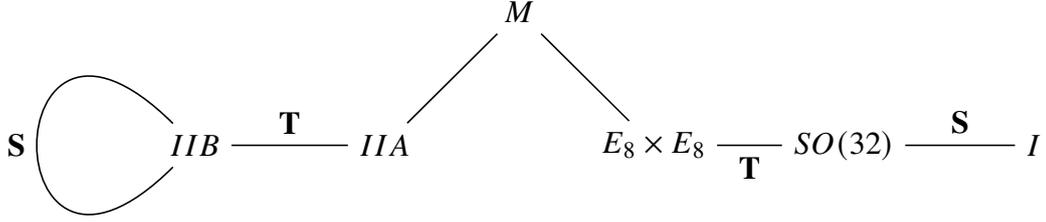

Taken together, these duality relations do not merely connect different regimes within a single string theory, but instead organize apparently distinct theories into a unified structure. The web of S- and T-duality implies that all five types of superstring theory (plus 11-dimensional M-theory, which is posited based on the analysis of the web of duality relations) are just different empirically equivalent formulations of the same theory (see Figure \ref{fig2}). 

\subsubsection{AdS/CFT correspondence}
 
 Moving on to AdS/CFT correspondence, we face a situation where only one part of the correspondence is a string theory at all. One starts with a string theory on anti-de Sitter space. In Maldacena's original proposal \citep{Maldacena:1997re}, this is $AdS_5\times S^5$, a $5$-dimensional anti-de Sitter space of radius $R$ with $5$ extra dimensions compactified on a 5-sphere, with the metric

\begin{equation}
ds^2 = R^2\frac{{\eta}_{\mu \nu} dx^{\mu}dx^{\nu} +dz^2}{z^2} +R^2d\Omega_5^2\,.   
\end{equation}

This theory is conjectured to be dual to a conformal $\mathcal{N}=4$ Super Yang-Mills theory on the 4-dimensional boundary of the stated AdS space. Once again, a near-classical description in one of the two is related to a deep quantum description in the dual representation. On AdS space, we find the near-classical theory in the limit where the AdS radius $R$ is large compared to the string length $l_s$. In this limit, gravitational effects are small and can, to a good approximation, be represented classically. In the dual superconformal Yang-Mills theory, the coupling strength is controlled by 
 \begin{equation}
\lambda:= g_{YM}^2N,  
 \end{equation}
where $N$ is the number of degrees of freedom of the Yang-Mills theory. Large $\lambda$ corresponds to a strongly coupled CFT. $\lambda$ is related to the string coupling on AdS by
\begin{equation}
g_s = \lambda / N .  
\end{equation}
Thus, in the large N limit a large $\lambda$ is compatible with a small string coupling $g_s$. On the other hand, $\lambda$ is related to the AdS radius by 

\begin{equation}
 \lambda \sim (R/l_s)^4 .  
\end{equation}
Therefore, large $\lambda$ corresponds to a large radius of AdS. A near-classical limit of the string theory on AdS is dual to a deep quantum description of CFT. 

AdS/CFT dualities have been conjectured in various scenarios and have been generalized beyond cases of strict conformal boundaries \citep{Hubeny_2015}. Speculations that a full gauge/gravity duality may apply to all string theoretical solutions have faced considerable obstacles, however, and remain elusive\footnote{For some problems that arise, see e.g. \cite{Goheer:2002vf}; for general issues that arise when trying to build meta-stable de Sitter solutions of string theory, see e.g. \cite{Obied:2018sgi}.}.

\subsubsection{The specific character of string dualities}

The case of string dualities differs from the context of dualities in quantum mechanics that have been analyzed above in two important ways.

First, the structural basis of the dualities is different. In quantum mechanics, the broad notion of duality refers to any isomorphism between two theories (Definition \ref{duality1}), while the more restrictive notion relevant for k-locality (Definition \ref{kduality}) presupposes a decomposition into subsystems.
In the context of string dualities, localization plays a pivotal but different role. All known dual string theories have small curvature, large compact radii, and low coupling constant limits where the spatial localization of objects is possible. Duality relations lead from such a near-classical limit in a given theory towards a dual theory that is not in a near-classical limit. The web of dualities, by establishing the empirical equivalence of theories with different classical limits, thus represents the set of classical limits of string theory. If string theory has a certain number of \textit{models} related by exact dualities, this implies that the theory has that number of classical limits. 

The second difference between string dualities and the previously discussed dualities in quantum mechanics is related to the first, but of a more pragmatic nature. To understand the point, we first need to say a few words about quantum field theory. A quantum field theory has a full formulation in terms of field operators and coupling constants. The full structure of the theory can be represented by a Lagrangian that characterizes the way fields propagate and couple to each other. To calculate the resulting dynamics, let us say in the case of a renormalizable gauge field theory, one needs to resort to perturbation theory. The latter provides an algorithm for calculating the theory's contributions to measurable amplitudes up to some order in coupling constants. Those calculations can be quite accurate if the coupling constants involved are small at the energy scale of the measured interaction process. Higher orders in perturbation theory are disregarded in the calculation. There are non-perturbative effects whose significance can be estimated roughly on a conceptual basis. For small coupling constants, those effects are understood to provide only small corrections to the perturbative results at low orders. In the case of a strongly interacting scenario, perturbation theory fails, but amplitudes can be estimated quantitatively based on various approximation methods. 

Even compared to the described quite non-trivial case of quantum field theory, string theory finds itself in a substantially more difficult situation. No full formulation of the theory is available. As described above, the conceptual starting point for spelling out the theory is perturbation theory for propagating and interacting strings. The tree level of perturbation theory, in conjunction with compact radii of a size that makes higher string winding modes heavy, allows for the identification of individual objects that behave in a near-classical way. As in quantum field theory, perturbation theory therefore points to a specific near-classical limit. Duality relations lead from the near-classical limit of one theory to a deep quantum state of a dual theory. 

All string dualities today have the status of conjectures, which are often supported by strong circumstantial evidence. As conjectures of exact dualities, they can only make sense, obviously, if the given dual theory, that is specified based on its perturbative expansion as well, has a well-defined deep quantum regime. The circumstantial evidence for the validity of duality relations thus suggests that the perturbative expansion of that theory identifies a well-defined full theory. Spelling out the web of dualities, therefore, is a way of representing the full theory that lacks a full formulation at this point.

\section{Common Core Realism and Related Views}\label{sec:common_core}

The examples of dualities discussed in the previous section raise an immediate challenge for traditional forms of scientific realism: if different theoretical formulations yield the same empirical predictions, what part of the theory, if any, deserves realist and ontological commitments? This pressure is not confined to string theory. Already in quantum mechanics, different TPSs on one and the same Hilbert space can render the very division into subsystems representation-dependent. Global isomorphisms that relate distinct TPS choices preserve all spectra and global expectation values while disagreeing about which degrees of freedom count as `local' and which interactions are $k$-local; in this precise sense, they provide dual representations of a single quantum system. Recent work has framed this as a \emph{quantum relativity of subsystems} \citep{AliAhmad:2021adn}, and related analyses emphasize that, given only $(\Hcal,\hat H,|\psi\rangle)$, there is in general no canonically preferred TPS unless additional relational data are supplied (cf. \cite{Stoica:2021rqi,SoulasFranzmannDiBiagio}). Taken at face value, these facts already undercut naive ontological realism about subsystem structure in QM and thereby mirror the string-theoretic erosion of a unique ontology.

Indeed, it has been suggested by a number of authors (\cite{Dawid2007, Rickles2011-RICAPL, Matsubara2013-MATRUA, Dawid2013, Rickles2017-RICDTS}) that the omnipresence of dualities in the context of string theory destroys the basis for realist commitment based on standard ontological realism. Further, it has been proposed in \cite{Dawid2007} and \cite{Dawid2013} to force structural realism to retreat towards a level of abstractness that reduces realist commitment to a holistic endorsement of a theory’s truth and fails to provide an interpretation in terms of individual elements of real structure.

A number of papers (\cite{Matsubara2018-MATSIS-4}, \cite{Rickles2017-RICDTS}, \cite{Huggett2017-HUGTS-2}), are guided by a more optimistic view on the prospects of a realist ontology of string theory. These lines of reasoning, which we will subsume under the name \textit{common core realism}, aim to salvage realism in the presence of dualities by identifying an invariant theoretical structure shared by all dual models. That \textit{bare theory} or \textit{common core} is suggested to provide a meaningful way to express realist commitment in light of duality relations. While those aspects of a dual representation that are not stable under duality transformation are unsuitable for realist commitment, what is common to all dual representations merits realist commitment as the \textit{common core} of the theory's \textit{models}. The proposal's basic idea is reminiscent of the distinction between physical and unphysical degrees of freedom in the context of gauge theory.

More recently, \cite{WuthrichHuggett2025} have focused their analysis on the issue of emergence, that is, on the question of how ontic commitments at an effective level are justified even if they do not reflect the fundamental ontology, while leaving open whether a common core of string theory can provide a sufficient basis for a realist ontology at a fundamental level. This approach resonates with the idea, familiar from \cite{Dennett1991-DENRP}'s notion of \emph{real patterns}, that ontological commitment can be grounded in the robustness and explanatory power of higher-level structures rather than in their reducibility to fundamental constituents. Similar ideas have recently been explored in physics by \cite{pittphilsci23307}, who argues that real patterns provide an appropriate ontological framework for emergent structures in physical theories. In this light, the emergentist realism of Wüthrich and Huggett can be seen as extending the pattern-based strategy to the domain of dualities, suggesting that even if no unique fundamental ontology is available, realist commitment may still attach to the stable, lawlike patterns that emerge across dual formulations.

A different angle on the role of the common core has been introduced by \cite{Read2020}. They argue that, in the absence of a common ontology, dual descriptions should not be taken to be \textit{physically equivalent}. That is, they should not be understood to represent the same physical theory. Expressed in common core terminology, they propose that, in the absence of a spelled-out common core that would be strong enough for implying a common ontology, no theory has been identified that is represented by both dual descriptions. The question as to what one should be realist about with regard to such a theory thus does not even arise. 

Read and M\o{}ller{-}Nielsen support their position, which they call the \textit{motivational approach} to symmetries and dualities, by analysing how the identification of a theory that comprises empirically equivalent formulations has been handled throughout the history of physics. Looking at examples such as the understanding of symmetries in the context of Newtonian physics, they argue convincingly that the acknowledgment of the existence of one theory with a certain symmetry structure is, in canonical physical examples, related to the attribution of a shared ontology to the representations that are linked by the symmetry. Based on this observation, they propose that an understanding of empirically equivalent descriptions in terms of one theory should, by definition, be conditional on identifying a common ontology of those descriptions. 

Applying their analysis to the case of string dualities, Read and M\o{}ller{-}Nielsen propose, in line with their general argument, that establishing the empirical equivalence of dual representations in a string theoretical context should not be taken to be sufficient for establishing that there is one \textit{string theory} that comprises those duals. Only the identification of a common ontology, which, on their view, might in the end be provided by M-theory, can establish the physical equivalence of the string theoretical duals and therefore provide the basis for talking about one overall \textit{string theory}.

The analysis of the common core has been made substantially more precise in the work of De Haro and Butterfield (\cite{DeHaro2018-DEHTHF}, \cite{Haro2019-HARTAT-43}, \cite{deHaroButterfield}), who provide an elaborate and formally worked out representation of the concept. On that basis, they respond to 
Read and M\o{}ller{-}Nielsen's reasoning. In a nutshell, de Haro and Butterfield argue that it is always possible to formally identify a common core of dual theories and, more importantly, that the common core theory can always be associated with a common core ontology. Therefore, they argue, the empirical equivalence of two dual representations that share the same empirical domain is sufficient for taking those representations to represent the same theory. Based on their discussion of the common core ontology,
De Haro and Butterfield propose what they call \textit{cautious realism} as the adequate form of realist commitment in the given context.

Given that theirs is the most extensive and formally precise take on the common core, we will, in the following, discuss the role of the common core and the prospects of a realism based on de Haro and Butterfield's view on the common core concept. In the next subsections, we introduce the formal apparatus of their approach and explore how it handles the relation between theory, model, and ontology. In Sec. \ref{sec:issuesCCR}, we present our criticism of common core realism on that basis. 

\subsection{De Haro and Butterfield on theories and models}

De Haro and Butterfield use slightly non-standard notions of ``model'' and ``theory''. 
On their account, models are specific realizations of a theory that, if related to each other by dualities, are empirically equivalent. The theory, then, on their account, is what the various dual models have in common. 

A model of a theory, therefore, consists of two parts, a \textit{root} part that is preserved across any isomorphic, dual transformation between different models, and a \textit{specific} part that is not preserved. It is the first part that they call the \textbf{common core theory}, or \textbf{bare theory}. To formulate a model of a common core theory, one needs to introduce \textit{specific structures}, which in most cases consist of a mathematical representation. 

More formally, de Haro and Butterfield define the bare theory in terms of a triple, consisting of: a space of states, $\mathcal{S}$, associated with different configurations, $s$, the system can be in; a set of physical quantities, $\mathcal{P}$, which takes values on the states; and a dynamics $\mathcal{D}$ which describes how these values change over time. Therefore, bare theories are schematically represented by the triple
\begin{equation}
    T := \langle \Scal, \mathcal{P}, \mathcal{D} \rangle\,.
\end{equation}
The value of a physical quantity $P\in \mathcal{P}$ in a state $s$ is represented as
\begin{equation}
    \langle P,s \rangle \,,
\end{equation}
and in quantum mechanics, for example, they correspond to the expectation value of any given operator: $\langle P,s \rangle = \bra{s} \hat{P} \ket{s} \in \mathbb{R}$.

The model, being a representation of the theory, is a \textit{homomorphic copy}\footnote{A homomorphism is a structure-preserving map between mathematical objects. In this context, a model is a homomorphic copy of the bare theory in the sense that it preserves the theoretical structure defined by the triple $\langle \mathcal{S}, \mathcal{P}, \mathcal{D} \rangle$, even if it introduces additional model-specific structure.}
 of the bare theory. Therefore, it also has a triple-like decomposition, however, loaded with specific structures. 
A model $M$ of a bare theory $T$ consists of its model root, $m = \langle \Scal_M, \mathcal{P}_M, \mathcal{D}_M \rangle$ plus its specific structure  $\Tilde{M}$, such that
\begin{equation}
    M:=\langle m, \Tilde{M}\rangle = \langle \Scal_M, \mathcal{P}_M, \mathcal{D}_M; \Tilde{M} \rangle\,.
\end{equation}
The \textbf{specific structures} comprise \textit{any parts and aspects of a model which do not express its realization of the bare theory}. 
 
Given that model roots are due to homomorphic maps of the bare theory, which do not necessarily imply injectivity or surjectivity, different realizations of the bare theory do not have to be isomorphic to it, nor isomorphic among themselves; but when they are, then we say they are \textit{dual}. 

One simple example to illustrate these concepts is to consider a single quantum particle in one dimension as a common core theory, where $\mathcal{S} = \Hcal (\mathbb{C}) \simeq L^2(\mathbb{R})$, the Hilbert space over the complex numbers; $\mathcal{P}=\{\hat{f}(\hat{X},\hat{P})\}$, meaning any function of the position and momentum operators, including the Hamiltonian $\hat{H}(\hat{X},\hat{P})$; and $\mathcal{D}$ is instantiated by the Heisenberg equation: $\dot{\hat{f}} = i [\hat{H},\hat{f}]$. This theory has two canonical representations in terms of the position and momentum bases, which are dual by a Fourier transformation, defining a unitary isomorphic map between the states and operators of each representation. Specifying a model then amounts to choosing a representation, i.e.\ a basis in which operators act explicitly. For example, in the position representation, states are given by wavefunctions $\psi(x) \in L^{2}(\mathbb{R})$, and the canonical operators act as $\hat{X} \psi(x) = x \psi(x)$ and $\hat{P} \psi(x) = -i \frac{d}{dx} \psi(x)$\footnote{These explicit forms for $\hat{X}$ and $\hat{P}$ apply in the Schrödinger picture, where operators are time-independent and act on time-evolving state vectors. In the Heisenberg picture, by contrast, the operators themselves evolve in time and are typically not represented in this static position-basis form.}. 

\subsection{The common core}

To introduce common core realism, we now return to the discussion of dualities in the context of theories and models defined in terms of their triples, and provide a more formal definition of dualities:

\begin{definition}[Duality]\label{duality3}
    A duality between models $m_i = \langle \mathcal{S}_{M_i}, \mathcal{P}_{M_i},\mathcal{D}_{M_i} \rangle$ requires: 
\begin{itemize}
    \item an isomorphism $d^{(i\rightarrow j)}_\mathcal{S}$ between the state-spaces, and 
    \item an isomorphism $d^{(i\rightarrow j)}_\mathcal{P}$ between the sets of physical quantities;
\end{itemize}
where $i,j$ label different models, such that the values of the physical quantities match, 
\begin{equation}
    \forall s_i \in \mathcal{S}_{M_i}\,, \forall \mathcal{P}_i \in \mathcal{P}_{M_i}\,, \text{ then } \langle \mathcal{P}_i,s_i \rangle_i = \langle d^{(i\rightarrow j)}_\mathcal{P}(\mathcal{P}_i),d^{(i\rightarrow j)}_\mathcal{S}(s_i) \rangle_j\,,
\end{equation}
and that the isomorphism maps are equivariant for the triples' dynamics, meaning that we can commute\footnote{This condition for commutation between the dynamical and duality maps is also introduced in the context of emergent maps in \citep{Carroll:2024nxc}.} either first evolving states and quantities in one model and then mapping them onto the other, or we can first map the states and quantities to another model and then consider their evolved dynamics\footnote{Not all theories take the form of a triple as specified here. Thus, one can broaden the definition of dualities to simply imply the existence of an isomorphism between models such that the values of the quantities remain the same under the map.}. 
\end{definition}

Notice that this definition does not say anything about how the domains of application of each model, $\mathcal{D}_{M_i}$, are related. In principle, each model has its own interpretation map, $i_{M_i}$, and they might not commute with the duality maps defined above, leading to \textit{physical/theoretical inequivalence} between the models; otherwise, if the interpretation and dual maps commute, then we say that the models are \textit{physically/theoretically equivalent}. \cite{deHaroButterfield} propose that for theoretically equivalent models, their ontologies are the same, as they describe exactly the same physics, and any difference between them is merely a difference of formulation, as their internal representations do not map these differences to anything in the world. 

Finally, we have all the needed elements to obtain the \textbf{common core theory}. The procedure of most interest to us is the following: 

\begin{quote}
    Suppose that we are given two dual models, $M_1$ and $M_2$. Then one way to obtain a (bare) common core theory $T$, that the two models represent or instantiate, is to ``cross out'' the structure of the models that is not common under the isomorphism. \citep[pg. 425]{deHaroButterfield}
\end{quote}
If one takes the two duals to be theoretically/physically equivalent, then the bare theory,  $T \cong m_1 \cong m_2$, has a \textit{unique internal representation map}, $i:T\rightarrow D$, defined in terms of the interpretation maps of the duals\footnote{De Haro and Butterfield also allow for a more extensive way of extracting a common core via \textit{augmentation}. In that case, the models $M_1$ and $M_2$ are first \textit{augmented} by adding more specific structure. The common core is then extracted from the augmented models. In their book, this strategy is used either to introduce a duality structure where the original models did not have one, or to establish the existence of a duality where the original models did not reveal it. Understood in this sense, augmentation seems not to be applicable in the cases of established or convincingly conjectured dualities in quantum mechanics and string theory that we are interested in. In private communication, de Haro and Butterfield indicated, however, that they aim to construe M-theory in terms of an augmentation of the 10-dimensional string theories rather than in terms of a dual model. While this step would render augmentation very relevant for the context we discuss, any in-depth analysis of this idea needs to wait until it has been spelled out.\label{augmentation}}.   

Given the above definition, one natural question is how far one should shave off a specific model such that what is left is isomorphic to another shaved-off model, thus allowing one to extract their common core. One example thoroughly discussed by \cite{deHaroButterfield} is bozonization, which takes two different classical models and shaves off their classicality, meaning the stipulation that the relevant quantized field either satisfies classical fermionic or bosonic equations of motion, leaving behind two fully quantum models that are dual. In their words, ``\textit{The two models are isomorphic to each other, in the sense that the bare theory is indeed their common core, but in addition each of the models has its own specific structure: namely, the classical structure that gives the operators (...) their specificity as (...) free bosons or as free chiral fermions.''}

As we will discuss, the bosonization example illustrates an extreme instance of shaving off specific structure, and is not representative of de Haro and Butterfield’s general doctrine. We believe that regarding classical structures or semi-classical limits as mere specific structures pushes the shaving procedure too far\footnote{Note that a field being bosonic or fermionic is a rather fundamental concept rooted in spacetime physics: the distinction is fixed by the field’s transformation properties under the Lorentz group, encoded in the Casimir operators of the Poincaré algebra. Treating such a structure as merely ``specific'' therefore amounts, in this context, to treating spacetime structure itself as specific.}. Before addressing this issue, let us now specialize the discussion of the common core in quantum mechanics and in string theory.

\subsection{The Common Core in Quantum Mechanics}\label{sec:CC_QM}

As we reviewed in Sec. \ref{sec:dualities}, dualities as defined in Definition \ref{duality3} are ubiquitous in quantum systems. In fact, this definition of duality only requires the existence of an isomorphic map while any single TPS is obtained as an equivalence class of these maps, as it mods out isomorphisms that are simply a product of local unitaries and permutations of quantum subsystems (see Definition \ref{def:TPS}).

Let's be a bit more explicit about this difference. Consider a single qubit. Any basis transformation, for instance from the computational basis to the $x$-basis, is an isomorphism satisfying all the properties of Definition \ref{duality3}, but it would be odd to consider that to be a duality transformation. The situation is analogous to taking the rotation of a coordinate system in Euclidean space to establish a duality. As we do not think of a basis transformation associated with a given physical system as a \textit{de facto} change of models, we will be conservative and retain our discussion at the level of transformations that change the TPS, since it is through the introduction of a TPS that we arrive at a notion of quantum subsystems.  

In any case, what is crucial for us is that any TPS transformation, therefore, is a duality transformation, and from the common core perspective, the very existence of dualities highlights specific structures that should be shaved off so that the common core is obtained. Explicitly, let's consider a quantum system $\mathcal{Q}$ which allows the introduction of multiple TPS $\mathcal{T}_i$, each defining a model $m_i = \langle \mathcal{S}_{M_i}, \mathcal{P}_{M_i},\mathcal{D}_{M_i} \rangle$. Each of these is composed of a model root and specific structures, and it's through the identification of the model root that we can arrive at the common core of these models. Given our discussion in Sec. \ref{sec:dualities_qm}, and the way we have introduced quantum systems, it is clear that we need to retreat to the global Hilbert space, its respective states, and observables acting on it, to define the common core in quantum mechanics,
\begin{equation}
    T= \langle \mathcal{H}, \mathcal{O}, H \rangle \,,
\end{equation}
where we represent the dynamical part of the triple by simply specifying the Hamiltonian $H$ of the system\footnote{In private correspondence, De Haro and Butterfield have emphasized that their framework does not, strictly speaking, \emph{require} TPSs to be “downstairs,” i.e., restricted to the level of models. In principle, a TPS could be included in the common core if it were preserved by all the relevant duality isomorphisms. However, this flexibility introduces a certain arbitrariness into the construction: the criterion for what belongs to the common core depends on which class of isomorphisms one takes to define the duality. Since the selection of those isomorphisms is guided by physicists’ judgments rather than by the formalism itself, the boundary between model-specific and invariant structure becomes pragmatically determined. To avoid this indeterminacy, we will assume that, in the context of quantum mechanics, different TPSs correspond to genuinely distinct models, and that no specific TPS structure should be part of the common core. We come back to this point in the Conclusion.}. 

According to common core realism, the introduction of a TPS constitutes the addition of specific structure—structure that varies across dual models and must therefore be excluded from the common core. But this move comes at a conceptual cost: TPSs are not merely representational choices. They determine the identification of subsystems, the structure of local observables, and the decomposition of the Hamiltonian into interaction terms. Most importantly, TPSs are what allow us to describe measurement contexts and formulate empirical predictions \citep{Zanardi:2004zz}. By relegating TPSs to the realm of specific structure, common core realism effectively strips away the very framework that connects the theory to observation. 

Once we let go of TPSs and retreat to the global Hilbert space and its associated dynamics and observables, then, from the perspective of common core realism, the only ontological content that survives across dual models is the structure defined over the global Hilbert: one is unavoidably concerned with the wavefunction of the whole Universe\footnote{Here we should note a crucial difference between the notion of duality à la \cite{Cotler:2017abq} and that of \cite{deHaroButterfield}. The former further imposes that the degree of locality in the Hamiltonian be preserved. Within that framework, one can still avoid dropping back to a purely global ontology by considering TPSs that, while inequivalent, preserve dynamical structure. The latter, by contrast, classifies all structure beyond the theory-defining triple as specific, thereby necessitating a return to global observables, states, and evolution.}. This idea, dating back to Everett's original formulation of the universal wavefunction \citep{Everett:1956dja,Everett:1957hd}, also appears in approaches to quantum cosmology (e.g., \cite{Bojowald:2015iga}) and more recently in proposals that treat the global Hilbert space as the bearer of fundamental ontology, such as \cite{Carroll:2021aiq}.

Whether this picture is satisfactory remains an open question. In Sec. \ref{sec:issues_CC_QM}, we argue that the appeal of common core realism in the quantum context overlooks essential aspects of how empirical observations are conducted, specifically, how they depend on the very structures that common core realism excludes from ontological consideration. In Sec. \ref{sec:proposal2}, we use this problematization to argue for a departure from Hilbert-space-based perspectives on realism, keeping it as the arena for ontology. 

\subsection{The Common Core in String Theory}\label{sec:CC_ST}

The extraction of a common core in string theory works along the same lines as in the context of quantum mechanics. The triple of states, quantities and dynamics is shared by all dual models and plays the role of the common core theory. The extra structure required for representing a specific type of string theory with its perturbative expansion, set of D-branes, etc, is not part of the common core and therefore does not enter the common core theory. Similarly, in the case of AdS/CFT correspondence, the correlation functions that provide the link between both sides of the duality enter the common core theory, but neither the stringy nature of the bulk theory nor the characteristics of a conformal field theory attributable to the boundary theory are part of the common core theory.
The common core theory extracted in the context of string dualities therefore is, in general, a highly reduced affair.

The situation becomes less clear once one moves from the common core theory to the common core ontology. While de Haro and Butterfield identify the common core theory by shaving off the extra structure of the dual models, they are willing to allow back in some elements of the dual ontologies as parts of the common core ontology via a process they call an augmentation of the dual models (see ft. \ref{augmentation}). They need this enrichment of the common core ontology to counter the charge that the common core ontology they extract, while formally existent, is too meager to be of any practical interest.

\subsection{Common Core Realism according to De Haro and Butterfield}\label{sec:CCR_dHB}

In Chapter 13 of their book, de Haro and Butterfield spell out the form of scientific realism they consider justified based on their discussion of the common core. They write: 

``...a cautious scientific realism notes that it is safe to commit
only to the weaker, more general, model that is empirically equivalent to, and
is fully compatible with, both duals, i.e. the common core theory, T.'' \citep[pg. 427]{deHaroButterfield}

The position they spell out in this quote does not insist that no realist commitment beyond the common core can ever be justified. Moreover, as described in the previous section, de Haro and Butterfield allow for the possibility that the common core ontology can be extended beyond what is strictly provided by shearing off all extra structure of individual models. Still, it is our understanding that their thinking does align with the basic principles of common core realism. Their basic tenet, as spelled out in the above quote, is the following: if a theory allows for more than one model, each fully spelled out model of the theory contains specific structures that may not merit realist commitment. A cautious approach to scientific realism, thus, should refrain from linking realist commitment to models but extract realist ontology from the theory that can be safely inferred to constitute the \textit{common core} of those models.

\section{Taking Issue with Common Core Realism}\label{sec:issuesCCR}

\subsection{The General Issue}\label{sec:general_issue}

The framework of common core realism offers a response to the problem of dualities: by identifying the common core shared across empirically equivalent models, it aims to preserve a prospect for realism while avoiding realist commitment to any particular dual formulation. 
As described in Sections \ref{sec:CC_ST} and \ref{sec:CCR_dHB}, de Haro and Butterfield aim to counter the charge of the insufficient richness of the common core ontology by allowing for its enrichment based on knowledge about dual models while still associating it with the common core theory. In the following, we will focus on the common core theory itself rather than the common core ontology and demonstrate that already the meagreness of the common core \textit{theory} generates very serious problems for realist commitment to that common core.

In both quantum mechanics and string theory, the structures that common core realism classifies as ``specific''—such as tensor product structures or near-classical limits—are precisely those that make empirical access to the theory possible.  We will argue that by excluding such structures from what one calls the \textit{theory}, one undermines the link between formalism and empirical content. What remains is a global structure that, while mathematically well-defined, cannot account for how we use these theories in practice. More specifically, the common core approach faces the following problem: 

\vspace{3mm}
\textit{In many important cases, the common core provides no basis for spelling out the empirical import of a scientific theory.}
\vspace{3mm}

To explain our point, we will rely on the notion of a \textit{compression algorithm} as introduced in \cite{Dennett1991-DENRP}. A compression algorithm substantially reduces the amount of information (bits, in Dennett's terminology) needed to fully characterize an observable system. Dennett takes the existence of a compression algorithm to indicate what he calls a ``real pattern'', a genuine regularity attributable to nature. We suggest that, on the canonical understanding of the concept of ``theory'', a theory amounts to a compression algorithm. 

A theory, as canonically understood, can be applied to a given interpreted data set and, on that basis, allows to extract predictions about the outcomes of other measurements. A classic example would be a theory that, applied to data characterising the state of the system at time $t_1$, predicts the state of the system at time $t_2$. The theory thus does two things: 

\begin{itemize}
    \item The theory connects to the collected data by giving it an interpretation;

    \item The theory amounts to a compression algorithm that substantially reduces the number of bits required for specifying the possible states of the system. 
\end{itemize} 

In de Haro and Butterfield's setup, it is \textit{models} in their sense that play the role of theories as canonically understood. They allow for extracting quantitative predictions when applied to initial conditions and therefore serve as compression algorithms. As we will elaborate in the context of quantum mechanics and string theory, in many important contexts, their \textit{common core theory} does not.

De Haro and Butterfield deploy the full triple as the sole basis for determining the bare theory because the latter is void of what they call the ``specific structures'' attributable to a model. The \textit{bare theory} is the full representation of states, quantities, and dynamics. It goes beyond a mere phenomenological representation of data since it relies on basic theoretical concepts such as states and quantities. But the definition of a triple does not require a framework for extracting states and quantities (spectra and observables) from the theory's defining structural principles. In other words, it does not require that the bare theory serve as a compression algorithm. 

As we argue, in some cases, it is indeed possible to enrich the triple in a way that turns it into a compression algorithm without relying on specific models. We encounter such cases both in quantum mechanics (the Fourier transform) and in string theory (self-dualities).
This feature, however, is not essential to de Haro and Butterfield's definition of a triple. In particular, it is absent in most dualities in string theory based on the current understanding of the theory; it may also be absent in more complex cases of duality relations in quantum mechanics.
In those cases, the common core lacks a crucial feature of a genuine scientific theory.

But why is it problematic to define \textit{theory} in a way that does not give it the role of a compression algorithm? Couldn't we view this step as an innocent matter of terminology? 
The serious problem arises once one uses this notion of theory as a basis for realist commitment. 

Let us briefly recall the role realist commitment played as a perspective on scientific theory. The understanding, expressed in the form of the so-called no-miracles argument \citep{putnam1975a,putnam1978}, was the following: the fact that a given theory is (approximately) true, and in some sense represents the real world, can explain why the theory's predictions are successful. This motivation for realist commitment, however, only works if fully stating that the true theory allows extracting the theory's predictions\footnote{One might be tempted to reinterpret the realist commitment discussed above within the framework of De Haro and Butterfield by simply mapping it onto their notion of a theory represented as a model, suggesting that it is the model, rather than the bare theory, that represents the real world and enables empirical success. However, this move is not available to them. On their ontological realist view, the real world is represented by the real ontology, and introducing realist commitment only at the level of specific models would undermine that stance. It would amount to accepting a form of mathematical Platonism where ontological significance attaches to invariant mathematical structures, independently of their empirical accessibility. Such a position, however, fails to explain how realism underwrites predictive success, and thus undermines the explanatory force of the no-miracles argument. This tension lies at the heart of the critique we develop in the following sections.}.  

In other words, de Haro's and Butterfield's account allows realist commitment at a level where the need for realism to explain predictive success is absent and refrains from realist commitment at the level (the level of \textit{models}) where predictive success plays out and would require an explanation. 

In the following, we will discuss how the general point discussed above plays out in the specific contexts of quantum mechanics and string theory.

\subsection{Quantum Mechanics} \label{sec:issues_CC_QM}

\begin{flushright}
    \small{\textit{``The emergence of a multi-partite tensor product structure of the state-space and the associated notion of quantum entanglement are (...) relative and observable-induced (...) by the experimentally accessible observables (interactions and measurements).''}

    \vspace{0.3em}
    -- \cite{Zanardi:2004zz}}
\end{flushright}

In the case of the duality between momentum and position representations in quantum mechanics, as spelled out above, it is possible to specify the equations of motion in a ``model''-independent way (simply by using abstract vectors in Hilbert space). Thereby, it is possible to understand the constraints on the dynamics in terms of the equations of motion without settling on an interpretation of the observables. We find a model-independent compression algorithm  because of the limited nature of the changes induced by the duality transformation (that is, the Fourier transformation.) 

As we argued in Sec. \ref{sec:CC_QM}, applying common core realism to quantum mechanics leads to a striking consequence: the real ontology of the theory is defined over the global Hilbert space, with no reference to subsystems, locality, or measurement. This is not merely a formal or representational shift, but it has deep implications for how the theory connects to the world. The structures that common core realism classifies as specific, such as tensor product structures, are precisely those that enable us to interpret the theory empirically. They define what counts as a subsystem, how observables are localized, and how measurement interactions are modeled. If these structures are excluded from the theory’s ontological content, it becomes unclear how the theory supports empirical predictions or explanatory practice at all.

Formally, this consequence can be made precise by identifying the theory’s invariant core with the triple $T = \langle \mathcal{H}, \mathcal{O}, \hat H \rangle$ defined over the global Hilbert space. Different tensor product structures, viewed as isomorphisms from $\mathcal{H}$ to factorized spaces whose factors represent subsystems, are then related by duality transformations and therefore count as specific rather than common structures. Since multiple inequivalent TPSs exist—each defining a distinct notion of subsystem and locality—they cannot all be part of the theory’s invariant content. The resulting picture is one of an essentially holistic ontology, as also emphasized by \cite{Carroll:2021aiq}, in which the theory describes the dynamics of the Universe as a single state vector in Hilbert space, with no privileged subsystem decomposition.

This raises a deeper concern: by excluding TPSs from the theory’s core structure, common core realism implicitly assumes that measurement and observation can be grounded in the global Hilbert space alone. But this is far from trivial. Modeling measurement without a subsystem decomposition requires resolving the measurement problem at the level of the universal wavefunction, a strategy associated with Everettian quantum mechanics, and even there, not without controversy. In practice, measurement involves interaction between subsystems, decoherence, and partial trace operations, all of which presuppose a TPS. Thus, common core realism sidesteps the measurement problem by assuming a kind of holistic closure that few working interpretations of quantum mechanics actually endorse.

We can further see how the common core perspective amounts to a form of holism by re-stating what has been coined as an \emph{intermediate notion of theory}. In \cite[pg. 30]{deHaroButterfield}, the state space of the theory is fixed (its dimensionality, for instance), so the theory cannot be seen canonically, such as Newtonian mechanics, where the state space varies depending on how many degrees of freedom the system has. Thus, their notion of intermediate theory keeps from the get-go a clear notion of a system, but that's exactly the notion that is put into question in the context of dualities, and more generally in the context of TPSs in quantum mechanics\footnote{Something similar also happens in string theory, where each different near-classical limit amounts to having different combinations of degrees of freedom, meaning different field content.}. For instance, different TPSs can break the state space of a quantum system in different sizes: more specifically, given $\mathcal{H} \overset{\Tcal}{\longrightarrow} \mathcal{H}_A \otimes  \mathcal{H}_B$, then $\text{dim } \mathcal{H} = \text{dim } \mathcal{H}_A \times \text{dim } \mathcal{H}_B$ is the only requirement, where $\text{dim } \mathcal{H}_A$ and $\text{dim } \mathcal{H}_B$ can vary depending on $\Tcal$. Hence, under the imposition of a fixed state space, and therefore fixed dimensionality of the Hilbert space, one is left only with the global Hilbert space to rely on, since the subsystems' state spaces are contingent on the TPS. 

However, is this satisfactory? Indeed, it is through a TPS that one can talk about the separation between system, apparatus, and environment, for instance, as well as the breakdown of the Hamiltonian into different components associated with different parts of the quantum system and their respective interactions, as well as their observables and states. It is this type of separation of the world into parts that guarantees us an empirical window into it. In other words, we do not have direct access to the world as a whole, but only to its parts. We argue that this hinders any reasonable notion of a fundamental holistic ontology in quantum mechanics, and poses a direct challenge to common core realism applied to quantum mechanics, while at the same time pointing out a way forward to reconcile our intuitive ideas about ontology with the commitment to the predictive import of the theory. 

The argument is based on the values that quantities in a given theory take upon a given state (see Eq. \eqref{eq:QM_empirical_import}). What are these values in practice? Quantum mechanically, following von Neumann's measurement theory, the overall quantum system composed of apparatus and physical system to be measured is represented as a factorization of the Hilbert space, $\Hcal = \Hcal_{\rm apparatus} \otimes \Hcal_{\rm system}$. A measurement is the evolution of the joint system such that the apparatus gets entangled with the measured system. Explicitly, let's assume for simplicity that both the apparatus and the measured system can be represented by qubits, then initially the apparatus is in the ``ready state'', $|r\rangle$, while the system might be in a superposition, say $\frac{1}{\sqrt{2}}\left(\ket{\uparrow}  + \ket{\downarrow} \right)$. Then, as apparatus and system get entangled, the initial product state becomes,
\begin{equation}
   |\Psi_0\rangle=\frac{1}{\sqrt{2}} \ket{r} \left(\ket{\uparrow}  + \ket{\downarrow} \right) \longrightarrow \ket{\Psi} = \frac{1}{\sqrt{2}} \left(\ket{\uparrow\uparrow} + \ket{\downarrow\downarrow} \right)\,.
\end{equation}
However, our empirical access is to the apparatus, not the measured system\footnote{We are being simplistic, as ultimately our access is to our phenomenal experience. That, though, can also be modeled. For a discussion, see \citep{DiBiagioFranzmann}.}. So, what we do is trace out the system's degrees of freedom. To do so, first we write the apparatus-system's density matrix, $\rho = |\Psi\rangle \langle \Psi |$, and find the apparatus' state by tracing out over the measured system,
\begin{equation}
    \rho_{\rm apparatus} = \text{Tr }_{\rm system} \ket{\Psi} \bra{\Psi}  = \frac{1}{2} \left(\ket{\uparrow} \bra{\uparrow} + \ket{\downarrow} \bra{\downarrow} \right)\,. 
\end{equation}
Finally, when we \emph{look} at the apparatus' display and see either spin up or down, thus the value $+1$ or $-1$, it is the value associated with the apparatus that we have access to, not the value associated with the combination of apparatus-system. Thus, the triple-breakdown of the theory fails to apply for the needed partition of the Hilbert space exactly when we need to have empirical access to the world, which is what instructs our ontological commitments. 

In other words, the only way that the values can result from a triple-like structure, where quantities are defined over the whole state space, is in the context of global observables and states, where nothing is ever traced out\footnote{Although see \citep{Franzmann:2024rzj} where one can connect the global observables to local ones by proposing a new solution to the measurement problem exactly by considering that quantum systems are not invariant general time evolution, i.e., the TPS is not preserved.}. However, in that context, we simply do not have a clear distinction between what measures and what is measured, as one could expect after the discussion regarding TPSs. Hence, the common core perspective loses touch with the theory's empirical import. 

Taken together, these considerations suggest that the common core approach, while elegant at the formal level, overlooks a crucial feature of how quantum mechanics connects to observations. If TPSs are necessary to define what counts as a subsystem, to model interactions, and to describe measurement outcomes, then their exclusion from the theory’s ontological content leaves that content disconnected from practice. Rather than reject realism altogether, this points to a need to rethink how ontology enters into our theories. Instead of seeking a fundamental ontology encoded in the global Hilbert space, we should treat ontology as emerging only within the contexts that enable empirical access. 

\subsection{String Theory}\label{sub:string_theory}

In the context of string theory, \textit{models} in the sense of de Haro and Butterfield are the individual duals of the theory. If we view the theory as a compression algorithm, at least one of the five types of string theory that are connected by duality relations needs to be spelled out to state the theory. The five types of string theory offer what comes closest to consistent and well-defined formulations of the theory. Moreover, they provide the basis for making predictions. To extract quantitative predictions of measurement outcomes, perturbative calculations of scattering amplitudes are carried out in one of five types of string theory.  

M-theory, which is related to 10-dimensional string theories by S-dualities, provides a deeper understanding of the overall theory and of the connectedness of the five types of 10 dimensional string theory. But, given that a quantum formulation of M-theory is not known at this point, a conceptualization of M-theory today needs to be anchored in the 10-dimensional string theories and 11-dimensional supergravity.

The question now is whether the common core of a set of duals can also provide a basis for understanding the theory's predictive power. For special constellations, the so-called \textit{self-dualities}, this is indeed the case. In self-dualities, the duality transformation links different parameter values within one and the same theory. In bosonic string theory, to give one example, T-duality implies that solutions where winding modes around a compact dimension are exchanged with transversal momenta along that dimension and the radius of the compact dimension is inverted are empirically equivalent. Solutions that seem entirely different when constructed classically are one and the same in a quantized system. The common core in this case comprises the full bosonic string theory and therefore does amount to a compression algorithm. The extra structure that characterizes the dual representations merely amounts to attributing different names to specific parameters based on the classical intuition that underlies the construction of a specific model. 
Similarly, type IIB superstring theory has an S-self-duality. Solutions of the theory with a given string coupling are empirically equivalent to solutions with an inverted string coupling. Type IIB string theory is the common core. 

But most string dualities are not self-dualities. They lead from one theory to a different one. A common core therefore must reduce to something narrower than any of the dual representations. Once one considers the full web of duality relations, entirely different representations enter the game: the five 10-dimensional superstring theories; 11-dimensional M-theory; Matrix theory as a structurally different representation of M-theory; a conformal field theory at the boundary of a string theory on AdS. Identifying the common core of that wide set of duals clearly could not identify anything resembling a theory that can serve as a compression algorithm. Rather, it would amount to a mere representation of data in terms of a record-keeping list of states and variables, as described in Sec. \ref{sec:general_issue}.

One way to avoid settling on the described minimal common core would be to reject the equality of all duals.
Indeed, there are certain elements of hierarchy between duals: M-theory is higher dimensional than the genuine string theories, it can represent the duality structure of the 10-dimensional theories, and it contains solutions that cannot be easily represented in 10 dimensions. The conformal theory on the boundary of AdS, in some respects, seems more straightforwardly calculable than the bulk theory. But in the end, provided string dualities are exact dualities, the duals still have the same observable spectrum and none of the duals is fully calculable. Any preference for one or the other ontology within the web of dualities thus seems to be based on subjective predilections for individual features or qualities of one or the other representation. 

One could still retreat to the position that a full understanding of string theory might reveal prospects for identifying a common core that can serve as a compression algorithm after all. This cannot be excluded, but it arguably should not be expected either.
A model-independent compression algorithm could be blocked at two levels. First, it might be the case that string theory does not have a set of fundamental equations of motion at all. In that case, the theory's implications would need to be, as a matter of principle, inferred based on consistency considerations starting from a partial specification of the theory, such as a perturbative expansion. In other words, the basic methodology applied in the context of string theory today would be all that could be applied in the context of string theory even in principle.

Second, one might imagine that there indeed are so far unconceived ways to spell out fundamental equations of motion for the individual ``models'' of string theory, but they lack a model-independent formulation. In particular when looking at the AdS/CFT correspondence, it is not implausible to think that there just is no possible conceptual physical framework that is strong enough to serve as a compression algorithm, and, at the same time, independent from the choice as to whether one wants to select a representation as a string theory or a representation as a conformal field theory. 

If either of the two described scenarios applies, string theory does not have a model-independent common core that can serve as a compression algorithm even in principle. We suggest that, under these conditions, no common core in string theory would be strong enough to provide the basis for a meaningful form of scientific realism. 

So far, we have argued that the common core of the duals in string theory currently is -- and may always remain -- too weak to serve as a basis for realist commitment. 
In the next Section, we will take this conclusion as the basis for proposing a different take on scientific realism in the presence of string dualities.
Before entering this discussion, however, it seems instructive to delve a bit deeper into the way our analysis affects the status of the common core in more general terms.

The reasoning presented above does not conflict with de Haro and Butterfield’s central argument that i) a common core can always be identified for a set of duals and ii) the common core's ontology safeguards that string duals satisfy Read and Møller-Nielsen’s criterion for physical equivalence. We agree, with the caveat that the common core theory may not function as a compression algorithm. Still, the general significance of the common core looks quite different in light of our analysis than from the perspective of de Haro and Butterfield or Read and Møller-Nielsen. As we will argue in the following, the specific nature of string theory does not fit well with using Read and M\o{}ller{-}Nielsen's criterion for physical equivalence.

String theory differs fundamentally from the earlier physical theories Read and M\o{}ller{-}Nielsen analyse in their paper.
As stated above, no fundamental equations of motion of string theory are known, which makes the relation between the theory's defining structural principles and its empirical implications notoriously difficult. While the five types of string theory are understood to represent a full-fledged theory, they are individually defined based on their perturbative expansions. Their non-perturbative import is to a large extent understood by making use of S-duality, which relates one theory in the deep quantum regime to its dual in a near-classical limit that allows for perturbative calculations.

The parameter values, such as compact radii or string couplings, which control whether a given type of string theory sits near a classical limit or not, are themselves determined by the theory's dynamics.  In other words, no formulation of the theory exists at this point that allows extracting the theory's empirical import directly from the theory's basic principles. 

In this light, there are strong reasons to view the connection between a theory’s fundamental structure and the spectrum of its representations differently in string theory than in other well-understood cases. While it seems plausible to say that one relies on a common core ontology when understanding, for example, the symmetry structure of Newtonian gravity as arising from a single underlying theory\footnote{The point is simply that Newtonian gravitation can be formulated in different but empirically equivalent ways, e.g. in terms of gravitational forces or in terms of inertial structure, and these formulations are naturally viewed as representations of one and the same theory because they share a clear, well-defined common ontology: absolute time, Euclidean space, inertial frames, and mass distributions. In such cases, identifying a common core ontology is both straightforward and explanatory.}, making the analogous claim in the context of string theory seems overly rigid. The absence of a neatly spelled out set of fundamental equations of the theory and the need to, in a sense, encircle the set of concepts that in conjunction represent the theoretical understanding of the physical situation, strongly suggest modesty when it comes to setting criteria for taking scientific approaches to be physically equivalent. This is in line with the way most string theorists think about the matter. In string physics, an exact duality relation is taken to be realized if the theories/models connected by the duality are empirically fully equivalent. If the two duals share the same empirical domain (which is trivially fulfilled in the case of a fully universal theory like string theory), one speaks about one theory with dual representations. No additional condition requiring a common ontology is needed. Any such additional condition would be considered an unnecessary further complication of a situation that is already difficult enough.

In the context of string dualities, focusing on ontology thus is unhelpful not only for understanding realist commitment but also for understanding physical equivalence. A conventional take on ontology just ceases to be helpful or instructive. It amounts to forcing a new scientific context into the mold of earlier periods of scientific reasoning that no longer fit.
To grasp the new situation on its own terms, one needs to fundamentally rethink the role of ontology in the context of fundamental physics.

\section{Separating Ontology from Scientific Realism}\label{sec:proposal1}

\subsection{The Basic Idea}

At first glance, one might take the discussion up to this point to suggest discarding the concept of ontology altogether as an outdated concept. This step, however, would fail to do justice to the significance of the observation that dualities do relate seemingly very different theories. It would also fail to explain why ontology was taken to be important for realist commitment before.

We suggest a different solution to the problem: view ontology and realist commitment as concepts that address two substantially different issues. 

\vspace{3mm}

{\bf Ontology} on our proposal is a characteristic of a near-classical limit, either tied to the specific structures (such as a tensor product structure) that enable a notion of locality, separability, and measurement in quantum mechanics, or to a particular perturbative expansion in string theory. Different dual models point at different such limits: in quantum mechanics, they instantiate different subsystem decompositions that ground empirically accessible observables; in string theory, they, in a way to be specified more clearly in the next subsection, point at different perturbative expansions around distinct vacua. In each case, the ontology arises from the structures that make the theory empirically meaningful in a given regime.

\vspace{3mm}

\textbf{Realist commitment}, by contrast, concerns the truth of the full theory. It targets the complete formal structure underlying all dual descriptions and therefore refrains from privileging the ontology of any single model. Ontology thus does not enter into the formulation of the theory’s fundamental content and plays no role in specifying what realist commitment requires.

The view we propose, then, is to decouple ontology from realism and instead treat it as a context-sensitive feature of regimes where empirical access is possible. We call this approach \textbf{observer-based ontologies}\footnote{As \cite{Dennett1991-DENRP} coined \textit{``real patterns''} to emphasize the reality of non-fundamental patterns, our terminology emphasizes the latching of ontology on experimental data.}: ontologies that are contingent and tied to specific representational structures, like subsystem decompositions in quantum mechanics or near-classical limits in string theory, but do not carry over to the level of fundamental description. Note that by ``observer-based'', we do not mean observation-dependent or perspective-dependent in Massimi’s sense \citep{Massimi2022-MASPRB}. What matters is not specific observations but the physical conditions that make empirical access possible.

Table \ref{tab:CCRvsESR} makes an explicit comparison between our proposal and common core realism.

\begin{table}[h!]
\renewcommand{\arraystretch}{1.5}
\setlength{\tabcolsep}{10pt}
\centering
\begin{tabular}{|p{3.cm}||>{\centering\arraybackslash}p{5cm}|>{\centering\arraybackslash}p{5.5cm}|}
\hline
\textbf{} & \textbf{Common Core Realism} & \textbf{Observer-based Ontology} \\
\hline\hline
\textbf{Realist \newline Commitment} & 
To the shared structure of dual models (the common core) & 
To the full theory as represented by models and duality transformations \\
\hline
\textbf{Ontology} & 
Extracted from what is common across duals & Linked to structures that point at possible empirical access; determined by a specific model. \\
\hline
\textbf{Role of Dualities} & 
Motivate search for unchanging structure; constrain ontology & 
Reveal observer-based ontology and motivate its separation from realism. \\
\hline
\textbf{Empirical Access} & 
Treated as external to the common core & 
Central: ontology arises only through structures like TPS or near-classical limits \\
\hline
\end{tabular}
\caption{Comparison between Common Core Realism and Observer-based Ontology.}
\label{tab:CCRvsESR}
\end{table}

In the following sections, we examine how the framework of realism and observer-based ontologies helps make sense of dualities while preserving a robust form of scientific realism, first in quantum mechanics and then in string theory.

\subsection{Empirical Accessibility and the Emergence of Ontology}\label{sec:emerge_onto}

A key feature of our proposal is that ontology is tied to empirical accessibility. By this, we mean the capacity of a theory to support observational or measurement-based claims, to connect its formal structure to phenomena. In both quantum mechanics and string theory, this capacity is not inherent to the theory’s common core but depends on specific structures. 

In quantum mechanics, empirical access dictates the specification of a TPS on the global Hilbert space \citep{Zanardi:2004zz}. This structure enables us to meaningfully distinguish between subsystems and their interactions, and it underwrites the operational framework of measurement: entanglement, decoherence, and the tracing out of subsystems are only definable relative to a TPS. Yet TPSs are not unique, and transformations between distinct TPSs (i.e., dualities) leave the global theory invariant. Thus, the ontology determined by measurement outcomes is not fundamental.

In string theory, dualities relate different near-classical limits of the theory, each of which in principle provides a basis for empirical access. But calculations of processes that play out in these near-classical limits cannot be based on the theory’s common core. They need to be carried out within the framework of the dual "model" that characterizes the near-classical limit and where the perturbation theory provides reliable quantitative predictions. Ontological claims, e.g., that the world consists of a certain type of strings and a specific set of D-branes and has a certain gauge symmetry, only make sense in a given near-classical limit. They are not licensed by the full theory, but by the specific ground state the observer happens to be in.

In both cases,  quantum mechanics as well as string theory, we find that empirical accessibility enables ontology, not the other way around. Ontology is not what grounds measurement or observation; but rather, it is what emerges from the structures that make measurement possible. This perspective is reinforced by the idea that ontology should reflect the presence of stable, compressible structure in the world, what Dennett and Wallace refer to as ``real patterns'' \citep{Dennett1991-DENRP,pittphilsci23307}. From this point of view, ontology is justified not by its fundamental status but by its explanatory power and empirical robustness.

\subsection{Realism without Ontology} \label{sec:proposal}

As we have argued, the framework of common core realism, while formally coherent, ultimately fails to connect realist commitment with the empirical content of physical theories. In both quantum mechanics and string theory, the structures that enable empirical predictions and measurements necessarily rely on features that are excluded from the common core. This observation motivates a different strategy: one that preserves the ambition of scientific realism, but detaches it from the demand for a real ontology. Instead, we propose a two-level picture: realism about the global structure of the theory, and observer-based ontology that emerges only in the context of empirical accessibility. In the next sections, we outline how this perspective plays out in the two main case studies of our paper.

Before turning to each case in detail, it is helpful to make the structural analogy between quantum mechanics and string theory more explicit (see Table \ref{tab:QMvsST}). While the two theories differ in scope and formulation, they exhibit striking parallels when it comes to the relationship between dualities, empirical access, and ontology. In both cases, we find that ontology is not grounded in the theory’s full invariant structure, but emerges only within particular regimes where the theory becomes empirically tractable. These regimes are defined by the specific structures that enable observation, measurement, and the near-classical limit. Dualities, rather than undermining this view, highlight its necessity by showing that multiple, equally valid ontologies can arise from different empirical perspectives on the same formal structure.

\begin{table}[h!]
\renewcommand{\arraystretch}{1.4}
\setlength{\tabcolsep}{10pt}
\centering
\begin{tabular}{|p{5.5cm}|p{5.5cm}|}
\hline
\textbf{Quantum Mechanics} & \textbf{String Theory} \\
\hline
\hline 
Tensor Product Structure specifies subsystems & Near-classical limit points at one of the duals. \\
\hline
Measurement depends on subsystem decomposition &  Near-classical limit/perturbative regime enables empirical access \\
\hline
Ontology emerges within TPS-defined contexts & Ontology emerges within near-classical limits \\
\hline
Global-Hilbert-space-based triple is the common  core & Invariant structures in the duality web define the common core \\
\hline
\end{tabular}
\caption{Analogy between quantum mechanics and string theory in the context of empirical access and ontology.}
\label{tab:QMvsST}
\end{table}

\section{Observer-based Ontology and Realism in Quantum Mechanics}\label{sec:proposal2}

\subsection{Observer-based Ontology}

In the context of quantum mechanics, the proposal of observer-based ontologies offers a natural response to the challenge posed by the multiplicity of inequivalent\footnote{See footnote \ref{ft:distinct_inequi}.} TPSs in a given Hilbert space. As reviewed in Sec. \ref{sec:dualities_qm}, a quantum system may be decomposed into subsystems in many different ways, each defined by an equivalence class of TPSs. These decompositions are not merely representational choices; they determine the system’s observable content, the structure of entanglement, and ultimately the conditions under which measurements can be described. 

In practice, however, not all TPSs are on equal footing. Our empirical standpoint as measuring agents effectively singles out those tensor-product structures that enable a meaningful link between the quantum formalism and the macrophysical world. Measurement interactions, decoherence, and the emergence of quasi-classicality all presuppose a decomposition of the Hilbert space that distinguishes accessible observables from environmental degrees of freedom. In this sense, empirical accessibility enforces a preferred TPS—one that is dynamically and operationally selected rather than arbitrarily imposed. This mirrors the situation in string theory, where empirical access similarly picks out specific near-classical limits among dual descriptions.

From the perspective of common core realism, TPSs are treated as ``specific structures'', optional additions to a core formalism that are stripped away when extracting a theory’s invariant content. The result is an ontology defined over the global Hilbert space alone, detached from any particular subsystem decomposition. As argued in Sec. \ref{sec:issues_CC_QM}, this view undercuts the very structures that make empirical access possible. Without a TPS, there is no clear distinction between what is measured and what does the measuring, no possibility of modeling decoherence, entanglement, or localized observables. In short, the common core lacks the conceptual machinery to account for how quantum mechanics connects to the world.

In contrast, the notion of observer-based ontologies acknowledges that ontological content in quantum mechanics only becomes meaningful once a TPS is specified. A TPS defines the operational interface between the theory and empirical practice: it tells us what counts as a system, what counts as an apparatus, and what observables are accessible \citep{Zanardi:2004zz,Stoica:2021rqi}. The ontology that arises within a given TPS, say, one in which the Hamiltonian is local and separable into interacting qubits, is not necessarily a reflection of a fundamental structure, but of the empirical context in which the system is probed. Different TPSs yield different ontological pictures (different decompositions of the Hamiltonian, different sets of local observables), and yet the underlying theory remains the same.

Crucially, these ontologies do not compete for real status. Rather, they are the only ontologies that make contact with the empirical world. This view also sheds new light on the role of decoherence, tracing out, and subsystem emergence. These processes require a TPS to even be formulated. The loss of coherence, the transition to classicality, and the appearance of stable pointer states are not features of the global Hilbert space in isolation but of its decomposition into parts. Hence, any ontology grounded in quantum measurement theory must be an observer-based ontology, one that depends on the very structures that common core realism dismisses.

In summary, quantum mechanics illustrates the need to separate ontological commitment from realism. The theory’s full formal structure admits multiple, inequivalent representations, none of which has a privileged claim to ``reality'' in the absence of empirical context. Ontology becomes meaningful only when tethered to the structures that render the theory testable and interpretable. Observer-based ontologies provide a framework for acknowledging this dependence while preserving the explanatory ambitions of realism.

\subsection{Realism}\label{subsec:realism_QM}

The case of quantum mechanics exemplifies the distinction between realist and ontological commitment. Realist commitment concerns the truth or approximate truth of the theory’s full formal structure—the Hilbert space, Hamiltonian, and algebra of observables—as a reliable framework for representing the world.  Ontological commitment, by contrast, concerns the entities or structures, such as a subsystem's decomposition given a TPS, that appear as elements of reality within specific empirical contexts. These two forms of commitment need not coincide. In what follows, we will see that while the Hilbert-space structure underlies ontology, the algebra of observables ultimately provides the more invariant locus for realism.

Although empirical access to a quantum system requires the introduction of a TPS on the global Hilbert space, such a structure is not uniquely fixed by the theory itself, which can typically be furnished by uncountably many different TPSs--all related by duality transformations. And that is exactly why, from the standpoint of common core realism, all possible TPSs count as ``specific structure'' and should therefore be excluded from the fundamental content of the theory. Meanwhile, from the standpoint of observer-based ontology, however, this structure is indispensable: it is precisely what allows ontology to emerge in the first place. 

In this sense, ontology in quantum mechanics is not part of the theory’s invariant core but arises within an empirical context, becoming unavoidably perspectival \citep{Adlam_2024}. We only ever interact with parts of the world, not with the global wavefunction, and those parts are specified by an empirical context that dictates the TPS we adopt in practice. The values of observables, and therefore any ontology we construct from them, depend on the ability to factorize the Hilbert space into empirically accessible components. Without that structure, the notions of measurement, locality, and classicality lose their footing, and with them, the very conditions under which ontology can be defined.

This echoes and offers a reframing of the classic observation by Landau and Lifshitz: 

\begin{quote}
    Thus quantum mechanics occupies a very unusual place among physical theories: it contains classical mechanics as a limiting case, yet at the same time it requires this limiting case for its own formulation. \citep{landau1991quantum}
\end{quote}
to which we could add: ``\textit{... for its own formulation at the empirical level.}'' Thus, within our new conceptual understanding, quantum mechanics still relates to claims of truth and reality, but it is its classical limit, obtained by classical measurements, that is responsible for its ontological structure: \textit{the world is quantum, while its ontology is near-classical.}

This dependence of ontology on classical limits and measurement contexts raises a further question: if the empirical interface fixes ontology, what remains as the invariant structure to which realism may attach?

\paragraph{Beyond Hilbert-space: an algebraic locus for realism.}

The discussion so far has emphasized that subsystem structure, and thus ontology, is contingent on empirical perspective. Yet one may ask whether a more invariant locus of realist commitment can be identified within quantum theory itself. The indeterminacy revealed by TPSs in Sec. \ref{sec:CC_QM} shows that subsystem structure, and thus the ontology inferred from it, is not fixed by the formal core of quantum theory. Which TPS we employ depends on operational or perspectival factors: the observables accessible to an agent, the partition of controllable degrees of freedom, or even the specification of a quantum reference frame. However, in all such cases, all the possible inequivalent TPSs are Hilbert-space representations of the same underlying algebraic structure. 

Therefore, to reconcile the TPS flexibility with the search for an invariant structure suitable for realism, one can shift the focus from Hilbert-space factorizations to the algebraic structure of observables itself. Working in the Heisenberg picture, we know that any unitary transformation--therefore any duality transformation--preserves the algebraic structure (see Appendix \ref{sec:descriptors}). On this view, TPSs are merely particular Hilbert-space representations realizing these algebraic relations.

This motivates a clean division of explanatory roles. Realism properly attaches to the algebraic structure, whereas ontology arises from a particular Hilbert-space instantiation of it. In this way, one retains a realist commitment to the algebraic structure common to all duals while allowing the ontology to vary with the TPS induced by observational perspective (as suggested by \cite{Zanardi:2004zz}). This departure from Hilbert-space-based realism in quantum mechanics aligns well with the situation in quantum field theory, where the difficulties in defining spatiotemporal local Hilbert spaces lead one towards algebraic quantum field theory, where local subalgebras are defined instead and preserved under spacetime symmetries.

\section{Observer-based Ontology and Realism in String Theory}\label{sec:new_ontology_ST}

We now turn to string theory, where the situation is in some ways more extreme than in the case of quantum mechanics: no full formulation of the theory exists, and all empirical access is mediated through perturbative dual models. As we will see, the framework of observer-based ontologies applies just as powerfully in that domain.

\subsection{Observer-based Ontology}
 
As described in Sec. \ref{sec:dualities_ST}, the types of string theory connected by dualities represent the same full theory but differ from each other with respect to their perturbative expansions. For weak couplings, the perturbative expansions point at the theory's near-classical limits and thereby at potential physical regimes in which measurement would be possible. If we carry out a measurement, we do so in one of the near-classical limits of the theory.

Let us understand more specifically, in the contexts of S- and T-duality, how the perturbative expansions single out a near-classical limit. In the case of S-duality, the situation is straightforward. S-duality connects a weakly-coupled theory to a strongly-coupled theory. In the weakly-coupled theory, the first orders of perturbation theory give fairly accurate results when calculating scattering amplitudes. In the strongly-coupled theory, to the contrary, the respective contributions to scattering amplitudes coming from higher orders in the coupling constant are not suppressed compared to lower orders, which means that the first orders in perturbation theory fail to give accurate results. At higher orders in perturbation theory, perturbation theory becomes unreliable, and non-perturbative effects become important. In other words, while an exact duality between the weakly- and the strongly-coupled theory (that is, including perturbative and non-perturbative effects) is conjectured, perturbative string theory is only applicable as a method for extracting scattering amplitudes with good accuracy in the case of the weakly-coupled theory. In this sense, the applicability of its perturbative theory singles out the type of string theory that is weakly coupled as the one pointing at the corresponding near-classical limit.

In the case of T-duality, one needs to take one extra step. If one starts in the weakly coupled regime, T-duality connects a weakly coupled type of superstring theory to another type of weakly coupled superstring theory with an inverted compactification radius and exchanged numbers of winding- and KK-modes. If the initial theory has a compact radius of the order string length, the energy gaps between KK-modes and transversal momenta are of comparable size. None of the duals allows for a near-classical description of the physics in the compact dimension since none shows the continuous spectrum of momenta that characterizes classical physics. Once we move towards a large radius limit in one of the T-duals, we approach a classical limit. The corresponding weakly-coupled theory thus characterizes a near-classical limit. Once we move towards a small radius limit, the KK modes approach a continuous spectrum. In order to understand this situation as a near-classical limit, however, we need to find a theory that represents this continuous limit as a spectrum of possible (classical) movements through space. We find this theory by going to the T-dual one, which exchanges winding-modes by transversal momenta and vice versa but, at the same time, also moves to a different type of superstring theory.  This T-dual type of string theory then represents the near-classical limit. 

In all of the described cases, the ontology associated with the perturbative expansion of the type of theory based on which a given ground state of the theory lies near a classical limit is the ontology attributable to an observer in that ground state. 
Each of the corresponding ontological statements derives its significance from the near-classical limit in which they can be represented in terms of human intuition: strings, in such a near-classical limit, can be understood as objects that can be localized and treated as nearly stable and separable from other objects. Curved spacetime can be understood as having sufficiently small curvature to allow for a smooth propagation of strings on it. 

\subsection{Realism}

Having thus established ontology as a characteristic of a given near-classical limit--and therefore of a given observer's perspective, we now move on to the next question: what is the status of the duality relations themselves?

The fact that dualities establish the empirical equivalence of the corresponding type of string theory with a different type of string theory, or an entirely different theoretical scheme that is not even a string theory narrowly construed, is not a statement about a given near-classical limit. Rather, it is a statement about the structural characteristics of the full theory. As discussed in Sec. \ref{sub:string_theory}, it is of eminent relevance for understanding the observational import of the theory. 

Realist commitment to string theory must amount to endorsing the truth or approximate truth of the overall theory, which should be expected to describe:

i) the fundamental mechanism that generates a world that contains a corner where we sit in some near-classical limit;

ii) the dynamics that play out in the near-classical limit that constitutes the part of the world we live in. 

Therefore, realist commitment must reach beyond our observer-perspective in explaining the dynamics that has led to its emergence. It is thus of essential importance for a realist understanding to be decoupled from our perspective, and, therefore, in light of the previous discussion, from a specific ontology. Realism about string theory, therefore, is neither about the common core -- which, as we have seen, is not capable of representing the theory's empirical import -- nor about any specific ontology that characterizes one of the duals. It is about a holistic endorsement of the entire set of statements about string theory that are expressed in a wide spectrum of dual representations, as well as the statements about the duality relations themselves.

Of course, a later stage in the understanding of string theory may render much of the present view unnecessarily complicated, maybe even misleading, but that does not invalidate realist commitment to the theory. Realist commitment about string theory asserts the truth of the full theory that is characterized and pointed at by the current statements of string theory, as incomplete as they may be.

\subsection{The Link to Effective Realism and Effective Ontology}\label{sec:link}

In order to fully understand the physical embedding of the split between realism and ontology, we need to think about it in terms of the tower of low-energy effective theories that lead from physics at low energies up to string theory. 

At the electroweak scale, the Standard Model of particle physics provides a fully accurate description of the phenomena. It is expected that, at higher energies, new degrees of freedom become empirically relevant and require a new theory. If string theory is a theory about nature, at the string scale, all degrees of freedom of string theory become visible. Looking at the situation from top to bottom, a more theory-driven view on effective theories focuses on steps of conceptualization rather than empirical data. On that account, 11-dimensional supergravity is an effective description of M-theory, while each of the five types of string theory has its 10-dimensional effective supergravity theory. Moving towards lower energies leads to 4-dimensional theories where the dualities as well as the supersymmetries are broken. Eventually, once again, one ends up at the standard model.

The standard model with its observed parameter values thus represents a low energy effective description of one ground state of string theory. It is based on assuming a four-dimensional background spacetime that has all the features we expect from a physical spacetime. Specifically, it allows for localization in a near-classical limit and for a continuous spectrum of momenta for freely moving particles. 

The question to ask about the spatiotemporal interpretation of superstring theories that are T-dual to each other is the following: given a certain radius of the extra dimension that is far from the string length, which of the parameters links up with a spatial parameter in the effective standard model description?    
The answer is straightforward. If one wants to identify the spatial dimensions of the fundamental theory with the spatial dimensions of a low-energy effective theory of the system, only the large compact dimension of the string theory can be called spatial.

Once one views the spatiotemporal ontology of string theory as a feature of its near-classical limit — linked to the ontology of the low-energy effective theory that emerges from that limit — the spatial parameter at the level of string theory becomes unequivocally identified.
The near-classical limit of the given dual would then be understood as the fundamental description of some low-energy effective field theory where the duality is broken and the ontology is uniquely identifiable as the ontology of the given theory. 

The described view connects naturally to work on \emph{effective scientific realism} \citep{doi:10.1093/bjps/axx043}. A dual model identifies one of string theory’s near-classical limits through its perturbative expansion; this limit determines the ontology appropriate to the type of perturbative string theory that can adequately describe that regime. Ontological commitment thus becomes indexed to patches of the landscape where measurement is physically meaningful. 
These patches serve as high-energy anchoring points for towers of effective theories where dualities no longer hold. In this sense, the ontologies associated with different duals are linked to \emph{effective} ontologies: emergent descriptions whose approximate truth is secured by the stability of the corresponding effective theory within its domain of applicability and which, in the absence of duality relations at their effective level, can be associated with a specific effective ontology. While realist commitment at the effective level thus can be associated with an ontology, realism at the more fundamental level of string theory, which pertains to the structure of the full theory that makes all effective descriptions possible, is decoupled from any ontological commitment.

\section{Conclusion}

In this paper, we found reasons for separating ontology from scientific realism in two important contexts of physical theory building. 

In the context of quantum mechanics, there is no fundamental reason to prefer a TPS over any other. All are fully adequate ways of viewing the system on an observer-independent basis. Only once we want to connect the ontology of our quantum system suggested by a specific TPS to our ontology of the macrophysical world, where our measurements play out, a preference for a specific TPS that allows for the localization of observables and a straightforward representation of decoherence emerges. Therefore, it is most natural to view ontology as a feature of an observer's perspective, whereas realism, qua being about the world independently of a human perspective, is about the theory of quantum mechanics without singling out an experimenter's perspective. Our proposal at that level is closely related to \cite{Adlam_2024}.

Now, at the level of quantum mechanics, one might ignore the conceptual separation between ontology and realism since one may expect, in real-world contexts, that there just is one TPS that allows an adequate representation of an observer-based perspective. Therefore, there is a sense in which the mere consideration that we need a measurable world to do physics at all already specifies the TPS.

Once one enters the regime of string dualities, however, this view becomes untenable. The theory now has, in view of its duality structure, several near-classical limits, each of them connected to different effective theories. If one viewed the landscape of solutions of string theory in the context of a multiverse theory, the different near-classical limits are not just possible outcomes of the dynamics of string theory. Rather, they are all descriptions of specific actualized corners of the multiverse. In other words, there are various actual observer perspectives, each of them linked to a different ontology; and there is one fundamental theory describing the physics that brings those perspectives about, which, if true, merits realist commitment. The separation of ontology from realism, therefore, becomes a necessary basis for understanding the peculiar relation between observer-independent theory and the observer-dependent connection of that theory to empirical data.

A related but conceptually deeper point emerges once one considers background-independent theories of quantum gravity. In quantum mechanics, the structures required for empirical access, such as subsystem decompositions, TPS choices, and measurement contexts, are externally fixed and not determined by the theory’s dynamics. In a background-independent setting \citep{read2024background}, no such structures can be posited; distinctions between systems, observables, or even spatiotemporal regions must arise endogenously from the theory itself. The very machinery that enables ontology in quantum mechanics must therefore emerge dynamically in quantum gravity. This strengthens the conclusion that realist commitment attaches to the invariant structure of the theory, while ontology can only appear as a near-classical, dynamical construct.

The striking parallels between the messages we get from quantum mechanics and then, once again, from string dualities go even further than just stated, however. In quantum mechanics as well as in string theory, we can understand why the identification of ontology and realism seemed initially like a natural idea. In the case of string theory, we have discussed this point already in Section \ref{sec:link}. String theory emerged from the paradigm of gauge field theory, which was deployed to low-energy phenomenology. In those effective gauge theories, however, the duality relations of fundamental string theory were broken. Therefore, it was possible to attribute one ontology to those theories, which allowed for linking realist commitment to an ontology.

A similar story can be told about quantum mechanics. 
Quantum mechanics has a classical limit. The classical theory, however, does not have a tensor product of state spaces, and the duality between different TPS does not arise. 
Therefore, from a classical perspective, identifying scientific realist commitment with a specific ontology looks natural.

Viewing ontology as a natural anchoring place for realist commitment was motivated by thinking about classical theorizing. In the contexts of quantum mechanics and, even more so, of string dualities, it is counterproductive and misleading and should be abandoned.

\section*{Acknowledgements}

RD and GF are thankful for many discussions with Jeremy Butterfield and Sebastian de Haro, for their careful reading of earlier versions of this manuscript, and for early access to their book on the physics and philosophy of dualities while this manuscript was being written. RF and GF are grateful for funding by the Swedish Research Council grant number 2022-01893\_VR. GF is thankful for financial support from the Olle Engkvist Foundation (no.225-0062)

\appendix

\section{Beyond the TPS} \label{sec:descriptors}

In the main text, we have largely framed the subsystem structure in quantum mechanics in terms of a tensor product structure (TPS) on the Hilbert space $\mathcal{H} = \mathcal{H}_A \otimes \mathcal{H}_B$. This decomposition enables the familiar picture of subsystems, local observables, entanglement, and measurement, and it grounds our account of \textit{observer-based ontologies} in empirically accessible degrees of freedom. However, this approach is limited in scope and ultimately inadequate for a fully general account of subsystems in quantum theory. 

A key problem is that \emph{locality with respect to a fixed TPS is not preserved under generic interacting time evolution}. In the Heisenberg picture, observables evolve via a unitary automorphism of the algebra $\mathcal{B}(\mathcal{H})$,
\begin{equation}
    \hat{\mathcal{O}}(t) = U^\dagger(t)\, \hat{\mathcal{O}} \, U(t),
\end{equation}
where $U(t) = e^{-i\hat H t}$ is generated by the total Hamiltonian $\hat H$. Even if the Hilbert space is initially written as $\mathcal{H}_A \otimes \mathcal{H}_B$, and $\hat{\mathcal{O}}_A \in \mathcal{B}(\mathcal{H}_A)$ is local at $t=0$, in general the evolved operator $\hat{\mathcal{O}}_A(t)$ will not be local with respect to the \emph{same} TPS at later times. This reflects the fact that interactions in $H$ generically generate entanglement (Schrödinger picture) or delocalize observables relative to the original factorization (Heisenberg picture).

It is important to stress, however, that \emph{unitary time evolution does not itself constitute a change of TPS}. The TPS remains a fixed kinematical structure on $\mathcal{H}$. What changes under time evolution is the embedding of observables into $\mathcal{B}(\mathcal{H})$, not the tensor factorization of the Hilbert space. While one may \emph{represent} the evolved observables as being local with respect to a \emph{different} TPS, this is a change of description rather than a physical transformation of subsystem structure.

The reason is that the \textit{algebraic relations} between observables are preserved under such dynamics. Suppose we begin with a von Neumann subalgebra $\mathcal{A} \subset \mathcal{B}(\mathcal{H})$ and its commutant $\mathcal{A}'$, with $\mathcal{A} \vee \mathcal{A}' = \mathcal{B}(\mathcal{H})$ (i.e. they generate the full algebra\footnote{A collection of subalgebras \(\{\mathcal{A}_\alpha\}_\alpha\) jointly generate all of \(\mathcal{B}(\mathcal{H})\),
\(
\bigvee_\alpha \mathcal{A}_\alpha \cong \mathcal{B}(\mathcal{H}),
\)
where \(\bigvee_\alpha \mathcal{A}_\alpha\) denotes the von Neumann algebra generated by the union of the \(\mathcal{A}_\alpha\), defined as the double commutant of the \(*\)-algebra generated by \(\cup_\alpha \mathcal{A}_\alpha\).}). This pair defines a notion of bipartition or subsystem structure at the algebraic level.

Time evolution induces an automorphism $\alpha_t$ of the full algebra,
\begin{equation}
    \alpha_t(\hat{\mathcal{O}}) := U^\dagger(t)\, \hat{\mathcal{O}} \, U(t),
\end{equation}
which maps $\mathcal{A}$ to another subalgebra $\alpha_t(\mathcal{A})$. Crucially,
\begin{equation}
    \alpha_t(\mathcal{A}') = \alpha_t(\mathcal{A})',
\end{equation}
so that the \emph{commutation relations, and hence the algebraic notion of subsystem independence, are preserved} under time evolution, even though the concrete realization of the subalgebras within $\mathcal{B}(\mathcal{H})$ changes. However, at each time the evolved pair $(\alpha_t(\mathcal{A}),\alpha_t(\mathcal{A})')$ defines a corresponding bipartition at the algebraic level; when additional conditions hold (mainly they are type I$_d$ von Neumann algebras), one can represent this bipartition by a unique Hilbert-space tensor factorization \citep{Zanardi:2004zz}. 

This point was first emphasized by \cite{Deutsch:1999jb} in the \emph{descriptor formalism} (see \citep{B_dard_2021} for a pedagogical review). More generally, algebraic relations between sets of observables are invariant under arbitrary unitary transformations, including those that would be interpreted, at the level of representation, as changing the TPS. This invariance motivates taking algebraic structure—rather than a particular tensor product decomposition—as the appropriate target of realist commitment, as argued in Sec.~\ref{subsec:realism_QM}.

\newpage 

\printbibliography

@article{Cotler:2017abq,
    author = "Cotler, Jordan S. and Penington, Geoffrey R. and Ranard, Daniel H.",
    title = "{Locality from the Spectrum}",
    eprint = "1702.06142",
    archivePrefix = "arXiv",
    primaryClass = "quant-ph",
    reportNumber = "SU-ITP-17-01",
    doi = "10.1007/s00220-019-03376-w",
    journal = "Commun. Math. Phys.",
    volume = "368",
    number = "3",
    pages = "1267--1296",
    year = "2019"
}

@article{Dawid2007,
	author = {Richard Dawid},
	journal = {Physics and Philosophy},
	title = {Scientific Realism in the Age of String Theory},
	year = {2007}
}

@article{Summers:2008zza,
    author = "Summers, Stephen J.",
    title = "{Subsystems and Independence in Relativistic Microscopic Physics}",
    eprint = "0812.1517",
    archivePrefix = "arXiv",
    primaryClass = "quant-ph",
    doi = "10.1016/j.shpsb.2009.02.002",
    journal = "Stud. Hist. Phil. Sci. B",
    volume = "40",
    pages = "133--141",
    year = "2009"
}

@book{Dawid2013,
	author = {Richard Dawid},
	editor = {},
	publisher = {Cambridge University Press},
	title = {String Theory and the Scientific Method},
	year = {2013}
}

@article{Matsubara2013-MATRUA,
	author = {Keizo Matsubara},
	doi = {10.1007/s11229-011-0041-3},
	journal = {Synthese},
	number = {3},
	pages = {471--489},
	publisher = {Springer Verlag},
	title = {Realism, Underdetermination and String Theory Dualities},
	volume = {190},
	year = {2013}
}

@article{Goheer:2002vf,
    author = "Goheer, Naureen and Kleban, Matthew and Susskind, Leonard",
    title = "{The Trouble with de Sitter space}",
    eprint = "hep-th/0212209",
    archivePrefix = "arXiv",
    doi = "10.1088/1126-6708/2003/07/056",
    journal = "JHEP",
    volume = "07",
    pages = "056",
    year = "2003"
}

@article{Obied:2018sgi,
    author = "Obied, Georges and Ooguri, Hirosi and Spodyneiko, Lev and Vafa, Cumrun",
    title = "{De Sitter Space and the Swampland}",
    eprint = "1806.08362",
    archivePrefix = "arXiv",
    primaryClass = "hep-th",
    reportNumber = "CALT-TH-2018-020, IPMU18-0100",
    month = "6",
    year = "2018"
}

@article{Huggett2017-HUGTS-2,
	author = {Nick Huggett},
	doi = {10.1016/j.shpsb.2015.08.007},
	journal = {Studies in History and Philosophy of Science Part B: Studies in History and Philosophy of Modern Physics},
	pages = {81--88},
	title = {Target Space $\neq$ Space},
	volume = {59},
	year = {2017}
}

@article{Matsubara2018-MATSIS-4,
	author = {Keizo Matsubara and Lars{-}G\"{o}ran Johansson},
	doi = {10.1007/s10838-018-9423-2},
	journal = {Journal for General Philosophy of Science / Zeitschrift f\"{u}r Allgemeine Wissenschaftstheorie},
	number = {3},
	pages = {333--353},
	publisher = {Springer Verlag},
	title = {Spacetime in String Theory: A Conceptual Clarification},
	volume = {49},
	year = {2018}
}

@book{Massimi2022-MASPRB,
	address = {New York},
	author = {Michela Massimi},
	editor = {},
	publisher = {Oxford University Press},
	title = {Perspectival Realism},
	year = {2022}
}

@article{Rickles2011-RICAPL,
	author = {Dean Rickles},
	doi = {10.1016/j.shpsb.2010.12.005},
	journal = {Studies in History and Philosophy of Science Part B: Studies in History and Philosophy of Modern Physics},
	number = {1},
	pages = {54--67},
	title = {A Philosopher Looks at String Dualities},
	volume = {42},
	year = {2011}
}

@article{Rickles2017-RICDTS,
	author = {Dean Rickles},
	doi = {10.1016/j.shpsb.2015.09.005},
	journal = {Studies in History and Philosophy of Science Part B: Studies in History and Philosophy of Modern Physics},
	pages = {62--67},
	publisher = {Elsevier Bv},
	title = {Dual Theories: ?Same but Different? or ?Different but Same??},
	volume = {59},
	year = {2017}
}

@article{DeHaro2018-DEHTHF,
	author = {De Haro, Sebastian},
	doi = {10.1007/s11229-018-1708-9},
	journal = {Synthese},
	pages = {1--35},
	title = {The Heuristic Function of Duality},
	year = {2018}
}

@article{Haro2019-HARTAT-43,
	author = {De Haro, Sebastian},
	doi = {10.1007/s13194-019-0261-9},
	journal = {European Journal for Philosophy of Science},
	number = {3},
	pages = {1--52},
	publisher = {Springer Verlag},
	title = {Towards a Theory of Emergence for the Physical Sciences},
	volume = {9},
	year = {2019}
}

@article{Stoica:2021rqi,
    author = "Stoica, Ovidiu Cristinel",
    title = "{3D-space and the preferred basis cannot uniquely emerge from the quantum structure}",
    eprint = "2102.08620",
    archivePrefix = "arXiv",
    primaryClass = "quant-ph",
    doi = "10.4310/ATMP.2022.v26.n10.a12",
    journal = "Adv. Theor. Math. Phys.",
    volume = "26",
    number = "10",
    pages = "3895--3962",
    year = "2022"
}

@article{Carroll:2021aiq,
    author = "Carroll, Sean M.",
    title = "{Reality as a Vector in Hilbert Space}",
    eprint = "2103.09780",
    archivePrefix = "arXiv",
    primaryClass = "quant-ph",
    reportNumber = "CALT-TH-2021-010",
    month = "3",
    year = "2021"
}

@book{landau1991quantum,
  title={Quantum Mechanics: Non-Relativistic Theory},
  author={Landau, L.D. and Lifshitz, E.M.},
  isbn={9780750635394},
  lccn={76018223},
  series={Course of theoretical physics},
  url={https://books.google.se/books?id=J9ui6KwC4mMC},
  year={1991},
  publisher={Butterworth-Heinemann}
}

@article{Adlam_2024,
   title={Moderate Physical Perspectivalism},
   ISSN={1539-767X},
   url={http://dx.doi.org/10.1017/psa.2024.73},
   DOI={10.1017/psa.2024.73},
   journal={Philosophy of Science},
   publisher={Cambridge University Press (CUP)},
   author={Adlam, Emily},
   year={2024},
   month=dec, pages={1–21} }

@article{EmerGe_proj_2,
  author       = {Colafranceschi, Eugenia and Di Biagio, Andrea and Flinckman, Joakim and Franzmann, Guilherme and Glowacki, Jan and Linnemann, Niels and Niedermann, Florian},
  collaboration = {EmerGe Collaboration},
  title        = {Subsystems Independence: from classical mechanics to quantum field theory and beyond},
  year         = {2026},
  eprint       = {in preparation},
  note         = {\url{https://emerge-collab.org}}
}

@phdthesis{Everett:1956dja,
    author = "Everett, III, Hugh",
    title = "{The Theory of the Universal Wave Function}",
    school = "Princeton U.",
    year = "1956"
}

@article{Everett:1957hd,
    author = "Everett, Hugh",
    title = "{Relative state formulation of quantum mechanics}",
    doi = "10.1103/RevModPhys.29.454",
    journal = "Rev. Mod. Phys.",
    volume = "29",
    pages = "454--462",
    year = "1957"
}

@article{Bojowald:2015iga,
    author = "Bojowald, Martin",
    title = "{Quantum cosmology: a review}",
    eprint = "1501.04899",
    archivePrefix = "arXiv",
    primaryClass = "gr-qc",
    doi = "10.1088/0034-4885/78/2/023901",
    journal = "Rept. Prog. Phys.",
    volume = "78",
    pages = "023901",
    year = "2015"
}

@article{Hubeny_2015,
   title={The AdS/CFT correspondence},
   volume={32},
   ISSN={1361-6382},
   url={http://dx.doi.org/10.1088/0264-9381/32/12/124010},
   DOI={10.1088/0264-9381/32/12/124010},
   number={12},
   journal={Classical and Quantum Gravity},
   publisher={IOP Publishing},
   author={Hubeny, Veronika E},
   year={2015},
   month=jun, pages={124010} }

@article{Maldacena:1997re,
    author = "Maldacena, Juan Martin",
    title = "{The Large $N$ limit of superconformal field theories and supergravity}",
    eprint = "hep-th/9711200",
    archivePrefix = "arXiv",
    reportNumber = "HUTP-97-A097, HUTP-98-A097",
    doi = "10.4310/ATMP.1998.v2.n2.a1",
    journal = "Adv. Theor. Math. Phys.",
    volume = "2",
    pages = "231--252",
    year = "1998"
}

@misc{polchinski2015dualitiesfieldsstrings,
      title={Dualities of Fields and Strings}, 
      author={Joseph Polchinski},
      year={2015},
      eprint={1412.5704},
      archivePrefix={arXiv},
      primaryClass={hep-th},
      url={https://arxiv.org/abs/1412.5704}, 
}

@article{Dennett1991-DENRP,
	author = {Daniel C. Dennett},
	doi = {10.2307/2027085},
	journal = {Journal of Philosophy},
	number = {1},
	pages = {27--51},
	publisher = {Journal of Philosophy Inc},
	title = {Real Patterns},
	volume = {88},
	year = {1991}
}

@article{Zanardi:2004zz,
    author = "Zanardi, Paolo and Lidar, Daniel A. and Lloyd, Seth",
    title = "{Quantum tensor product structures are observable induced}",
    eprint = "quant-ph/0308043",
    archivePrefix = "arXiv",
    doi = "10.1103/PhysRevLett.92.060402",
    journal = "Phys. Rev. Lett.",
    volume = "92",
    pages = "060402",
    year = "2004"
}

@misc{pittphilsci23307,
           month = {April},
           title = {Real Patterns in Physics and Beyond},
          author = {David Wallace},
            year = {2024},
             url = {https://philsci-archive.pitt.edu/23307/},
}

@article{AliAhmad:2021adn,
    author = "Ali Ahmad, Shadi and Galley, Thomas D. and Hoehn, Philipp A. and Lock, Maximilian P. E. and Smith, Alexander R. H.",
    title = "{Quantum Relativity of Subsystems}",
    eprint = "2103.01232",
    archivePrefix = "arXiv",
    primaryClass = "quant-ph",
    doi = "10.1103/PhysRevLett.128.170401",
    journal = "Phys. Rev. Lett.",
    volume = "128",
    number = "17",
    pages = "170401",
    year = "2022"
}

@article{Franzmann:2024rzj,
    author = "Franzmann, Guilherme",
    title = "{To be or not to be, but where?}",
    eprint = "2405.21031",
    archivePrefix = "arXiv",
    primaryClass = "quant-ph",
    month = "5",
    year = "2024"
}

@book{putnam1975a,
  author    = {Hilary Putnam},
  title     = {Mathematics, Matter and Method},
  year      = {1975},
  volume    = {1},
  series    = {Philosophical Papers},
  publisher = {Cambridge University Press},
  address   = {Cambridge}
}

@incollection{putnam1978,
  author    = {Hilary Putnam},
  title     = {Realism and Reason},
  booktitle = {Meaning and the Moral Sciences},
  pages     = {123--138},
  year      = {1978},
  publisher = {Routledge and Kegan Paul},
  address   = {London}
}

@article{Carroll:2024nxc,
    author = "Carroll, Sean M. and Parola, Achyuth",
    title = "{What Emergence Can Possibly Mean}",
    eprint = "2410.15468",
    archivePrefix = "arXiv",
    primaryClass = "physics.hist-ph",
    month = "9",
    year = "2024"
}

@book{WuthrichHuggett2025,
  author    = {Christian W\"uthrich and Nick Huggett},
  title     = {Out of Nowhere: The Emergence of Spacetime in Theories of Quantum Gravity},
  publisher = {Oxford University Press},
  year      = {2025},
  address   = {Oxford},
  pages     = {304},
  isbn      = {9780191076152},
}

@article{Read2020,
	author = {James Read and Thomas M\o{}ller{-}Nielsen},
	journal = {Synthese},
	number = {1},
	pages = {263--291},
	publisher = {Springer Verlag},
	title = {Motivating Dualities},
	volume = {197},
	year = {2020}
}

@book{deHaroButterfield,
  author       = {De Haro, Sebastian and Jeremy Butterfield},
  title        = {The Philosophy and Physics of Duality},
  publisher    = {Oxford University Press},
  year         = {2025},
  address      = {Oxford},
  isbn         = {978-0198846338},
  pages        = {624},
  note         = {Open access under CC BY‑NC‑ND 4.0},
}

@article{Deutsch:1999jb,
    author = "Deutsch, David and Hayden, Patrick",
    title = "{Information flow in entangled quantum systems}",
    eprint = "quant-ph/9906007",
    archivePrefix = "arXiv",
    doi = "10.1098/rspa.2000.0585",
    journal = "Proc. Roy. Soc. Lond. A",
    volume = "456",
    pages = "1759--1774",
    year = "2000"
}

@article{B_dard_2021,
   title={The ABC of Deutsch–Hayden Descriptors},
   volume={3},
   ISSN={2624-960X},
   url={http://dx.doi.org/10.3390/quantum3020017},
   DOI={10.3390/quantum3020017},
   number={2},
   journal={Quantum Reports},
   publisher={MDPI AG},
   author={Bédard, Charles},
   year={2021},
   month=apr, pages={272–285} }

@unpublished{SoulasFranzmannDiBiagio,
    author = "Soulas, Antoine and Franzmann, Guilherme and Di Biagio, Andrea",
    title = "{On the emergence of preferred structures in quantum theory}",
    eprint = "2512.07468",
    archivePrefix = "arXiv",
    primaryClass = "quant-ph",
    month = "12",
    year = "2025"
}

@unpublished{DiBiagioFranzmann,
  author  = {Andrea Di Biagio and Guilherme Franzmann},
  title   = {A multiversal double-standard?},
  year    = {2025},
  note    = {In preparation}
}

@book{read2024background,
  title={Background Independence in Classical and Quantum Gravity},
  author={Read, J.},
  isbn={9780192889119},
  url={https://books.google.se/books?id=RxzcEAAAQBAJ},
  year={2024},
  publisher={Oxford University Press}
}

@article{doi:10.1093/bjps/axx043,
author = {Williams, Porter},
title = {Scientific Realism Made Effective},
journal = {The British Journal for the Philosophy of Science},
volume = {70},
number = {1},
pages = {209-237},
year = {2019},
doi = {10.1093/bjps/axx043},

URL = {https://doi.org/10.1093/bjps/axx043},
eprint = {https://doi.org/10.1093/bjps/axx043}
}

\end{document}